\documentclass{aa}
\usepackage{graphicx}
\usepackage{float}
\graphicspath{ {./plots/} }
\usepackage{txfonts}
\usepackage{natbib}
\bibpunct{(}{)}{;}{a}{}{,}
\usepackage{multicol}
\usepackage{array,multirow}
\usepackage{footnote}
\usepackage{url}
\usepackage[titletoc]{appendix}
\usepackage{cleveref}
\usepackage{soul}
\usepackage[switch]{lineno}
\begin{document}

   \title{Localizing the $\gamma$-ray emitting region in the blazar TXS~2013$+$370}

%   \subtitle{The location of GeV emission production}

    \author{E. Traianou
          \inst{1} \thanks{Member of the International Max Planck Research School for Astronomy and Astrophysics at the Universities of Bonn and Cologne.}
          \and
          T.~P.~Krichbaum\inst{1}\fnmsep\ B.~Boccardi \inst{1,2} \and R.~Angioni\inst{1,3}\thanks{Now at Space Science Data Center – Agenzia Spaziale Italiana, Via del Politecnico, snc, I-00133 Roma, Italy and INFN - Roma Tor Vergata Via della Ricerca Scientifica, 1. I-00133 Rome, Italy} \and B.~Rani \inst{4} \and J.~Liu \inst{1} \and E.~Ros\inst{1} \and U.~Bach\inst{1} \and K.~V.~Sokolovsky\inst{9,10,11}   \and S.~Kiehlmann\inst{5,6,7} \and M.~Gurwell\inst{8} \and J.~A.~Zensus\inst{1}
          }

   \institute{Max-Planck-Institut f\"ur Radioastronomie, Auf dem H\"ugel 69, D-53121, Bonn, Germany\\
              \email{etraianou@mpifr.de}
             \and
             INAF -- Osservatorio di Astrofisica e Scienza dello Spazio di Bologna, Via Gobetti 101, I-40129 Bologna, Italy 
             \and
   Institut f\"ur Theoretische Physik und Astrophysik, Universit\"at
   W\"urzburg, Emil-Fischer-Str. 31, D-97074 W\"urzburg, Germany
             \and
             NASA Goddard Space Flight Center, MD 20771, Greenbelt, USA
             \and 
             Institute of Astrophysics, Foundation for Research and Technology-Hellas, GR-71110 Heraklion,Greece
             \and
             Department of Physics, Univ. of Crete, GR-70013 Heraklion, Greece
             \and
             Owens Valley Radio Observatory, California Institute of Technology, CA 91125, Pasadena, USA
             \and 
             Center for Astrophysics ~$\mid$ Harvard \& Smithsonian, 60 Garden St, MA 02138, Cambridge, USA
             \and
             Department of Physics and Astronomy, Michigan State University, East Lansing, Michigan 48824, USA
             \and
             Sternberg Astronomical Institute, Moscow State University, Universitetskii pr. 13, 119992, Moscow, Russia
             \and
             Astro Space Center of Lebedev Physical Institute, Profsoyuznaya St. 84/32, 117997, Moscow, Russia
             }

   \date{Received ???????; accepted ?????}

% \abstract{}{}{}{}{} 
% 5 {} token are mandatory

  \abstract
  {} % leave it empty if necessary  
   {The $\gamma$-ray production mechanism and its localization in blazars are still a matter of debate. The main goal of this paper is to constrain the location of the high-energy emission in the blazar TXS~2013$+$370 and to study the physical and geometrical properties of the inner jet region on sub-pc scales.
   } 
   {
   TXS~2013$+$370 was monitored during 2002-2013 with VLBI at 15, 22, 43, and 86\,GHz, which allowed us to image the jet base with an angular resolution of $\geq$ 0.4\,pc.
   By employing CLEAN imaging and Gaussian model-fitting, we performed a thorough kinematic analysis at multiple frequencies, which provided estimates of the jet speed, orientation, and component ejection times.
   Additionally, we studied the jet expansion profile and used the information on the jet geometry to estimate the location of the jet apex. VLBI data were combined with single-dish measurements to search for correlated activity between the radio, mm, and $\gamma$-ray emission. For this purpose, we employed a cross-correlation analysis, supported by several significance tests.
   }  
   %results heading (mandatory)
   {
   The high-resolution VLBI imaging revealed the existence of a spatially bent jet, described by co-existing moving emission features and stationary features. New jet features, labeled as A1, N, and N1, are observed to emerge from the core, accompanied by flaring activity in radio/mm- bands and $\gamma$ rays. The analysis of the transverse jet width profile constrains the location of the mm core to lie $\leq$2\,pc downstream of the jet apex, and also reveals the existence of a transition from parabolic to conical jet expansion at a distance of ${\sim}54$\,pc from the core, corresponding to ${\sim}1.5\times10^{6}$ Schwarzschild radii. The cross-correlation analysis of the broad-band variability reveals a strong correlation between the radio-mm and $\gamma$-ray data, with the 1\,mm emission lagging $\sim 49$\,days behind the $\gamma$ rays. Based on this, we infer that the high energy emission is produced at a distance of the order of $\sim$1\,pc from the jet apex, suggesting that the seed photon fields for the external Compton mechanism originate either in the dusty torus or in the broad-line region.} 
   {}
   \keywords{VLBI --
                jet --
                blazar
               }

\titlerunning{TXS~2013$+$370: A $\gamma$-ray loud blazar at ultra high angular resolution}
\authorrunning{E. Traianou et.al.} 

\maketitle
%
%________________________________________________________________

\section{Introduction}
\label{sec:intro}
%\linenumbers

Matter accretion onto a supermassive black hole (SMBH) -- one of the most efficient energy production mechanisms in the Universe \citep{2011ApJ...728...98D} -- is thought to power Active Galactic Nuclei (AGN). 
This physical process, combined with the presence of strong magnetic fields accumulated in the accretion disk, is linked to the formation of collimated plasma jets emanating from the nuclear regions and propagating up to large distances \citep{1977MNRAS.179..433B, 1982MNRAS.199..883B}. 
Powerful AGN jets, especially those oriented close to our line of sight (i.e., blazars), are observed to radiate over the entire electromagnetic spectrum and to display extreme variability \citep[e.g.,][]{Aharonian_2007}. 
A persistent and intriguing question, specifically relevant to the high-energy part of the emission, concerns the physical processes driving the jet $\gamma$-ray emission and the location of its production site.
Observational findings and theoretical models suggest that the high energy emission can be produced in regions close to the central engine as well as further downstream in the jet \citep[e.g.,][and references therein]{2001ApJ...556..738J, 2014A&A...571L...2R, review}.

In the framework of leptonic models, the broadband AGN emission is thought to be produced by leptons ($e^-$ and $e^+$) through synchrotron and inverse Compton (IC) processes. The photons that are scattered up to $\gamma$-ray energies can either be the same synchrotron photons radiated by the jet (Synchrotron-Self-Compton, \citealt{1992ApJ...397L...5M}) or can originate in the jet surroundings 
(External Compton) \citep[e.g.,][]{2008MNRAS.386L..28G,2009ApJ...692...32D}. 
With increasing distance from the black hole, possible reservoirs of these seed photons are the accretion disk  \citep{1992A&A...256L..27D}, the broad-line region \citep[BLR,][]{1993ApJ...416..458D,1994ApJ...421..153S,2010ApJ...717L.118P,2012ApJ...758L..15D}, or the dusty torus \citep{2000ApJ...545..107B,1999ApJ...514..138K}. At larger distances, Cosmic Microwave Background (CMB) photons may play a role \citep{2008MNRAS.385..283C}. If ultra-relativistic protons, as well as leptons, compose the jet, $\gamma$ rays could also be produced through proton-synchrotron or photo-pion production \citep{1992A&A...253L..21M, 2000NewA....5..377A}.

Very-long-baseline interferometry (VLBI) observations at millimeter wavelengths are ideally suited to investigate the origin of $\gamma$-ray emission in blazars as they provide a very sharp view of the innermost jet regions, generally affected by synchrotron opacity effects at longer wavelengths. When used in combination with monitoring of the broadband emission variability, mm-VLBI can enable us to effectively pinpoint the location of the $\gamma$-ray production site \citep[see][and references therein]{2017A&ARv..25....4B}. 

In this study, we focus on the compact radio source TXS~2013$+$370, which is associated with a $\gamma$-ray loud object \citep{2000ApJ...542..740M,2001ApJ...551.1016H,2012ApJ...746..159K,2016A&A...594A..60L} of an uncertain type \citep{2015Ap&SS.357...75M} 
at redshift $z = 0.859$ \citep{2013ApJ...764..135S} and hosting a SMBH with a~mass of $4\times10^{8}$\,M$_\sun$ \citep{2015MNRAS.448.1060G}. 
Based on ultra-high-resolution VLBI and single-dish observations performed over 10 years, we obtained a detailed description of the morphological evolution and the variability properties of the radio jet. By combining a kinematic and geometrical analysis of the jet base with the investigation of correlated flux density variability in the cm- and mm- radio and in the $\gamma$-ray bands, we were able to constrain the location of the $\gamma$-ray production region.

The article is organized as follows. In Sect.~\ref{sec:obs/im}, we present the multi-frequency data set and the data reduction techniques; in Sect.~\ref{sec:analysis_results}, we report on the results from the VLBI study and the multi-band variability analysis; in Sect.~\ref{sec:discussion}, we discuss their implications for the $\gamma$-ray production; in Sect.~\ref{sec:conclusions}, we summarize our conclusions.  
For our calculations, we adopt the following cosmological parameters: $\Omega_\mathrm{M}=0.27$, $\Omega_{\Lambda}=0.73$, $\rm H_{0}=71$\,km~s$^{-1}$~Mpc$^{-1}$ (similar to those used by \citealt[and references therein]{2016AJ....152...12L}), which result in a luminosity distance $D_{L}=5.489$\,Gpc and a linear-to-angular size conversion of 7.7\,pc/mas for the redshift of z=0.859.

\section{Observations, data calibration and imaging}
\label{sec:obs/im}

\subsection{VLBI observations}
\label{subs:vlbi_obs}

Our VLBI data set includes observations at 15, 22, 43, and 86\,GHz. 
The details of the VLBI observations are summarized in Table~\ref{table:datalog}.

The 86\,GHz observations were performed with the Global Millimeter VLBI Array (GMVA\footnote{\url{https://www3.mpifr-bonn.mpg.de/div/vlbi/globalmm/}}) and the 43\,GHz observations with the Global VLBI array; the four epochs were observed in 2007--2009 and three epochs in 2009--2010, respectively.

For a single epoch, the {\em RadioAstron} space antenna \citep{2013ARep...57..153K}, in combination with the VLBI ground array performed simultaneous observations at two frequencies, 22\,GHz and 5\,GHz, to facilitate fringe search at the space-ground baselines. In this article we consider the 22\,GHz data.

The space-VLBI data were correlated with a special {\em RadioAstron}-enabled version 
\citep{2016Galax...4...55B} of {\tt DiFX} software 
correlator running on a desktop computer. The fringe search performed with {\tt PIMA} \citep{2011AJ....142...35P} resulted in space-ground fringe detection at baselines up to 1.7~Earth diameters. The full description of the space-VLBI experiment will be given by K. Sokolovsky~et.~al. (in~prep.). At 22\,GHz, we re-imaged the source based on the data originally presented by 
\cite{2014cosp...40E3161S} and \cite{2015SoSyR..49..573K}, who also show a 5\,GHz image. 

At 15\,GHz we re-analyzed fifteen epochs of Very Long Baseline Array \citep[VLBA,][]{1994IAUS..158..117N} observations that cover a period from 2002 to 2012 and are publicly available at the MOJAVE data archive\footnote{\url{https://www.physics.purdue.edu/MOJAVE/sourcepages/2013+370.shtml}} 
\citep[see][and references therein]{2009AJ....137.3718L,2011ApJ...742...27L,2018ApJS..234...12L}. 

\subsubsection{VLBI data calibration}
\label{subs:vlbi_data_calb}

The data reduction was performed using the National Radio Astronomy Observatory's (NRAO) Astronomical Image Processing System \citep[AIPS,][]{1990apaa.conf..125G}. 
The calibration of the GMVA data at 86 and 43\,GHz was performed in the standard manner \citep[e.g.,][]{2019A&A...622A..92N}: 
after an initial parallactic angle correction of the phases, we determined the inter-band phase and delay offsets between the intermediate frequencies using some high signal-to-noise ratio scans (manual phase calibration, \citealt{2012A&A...542A.107M}). 
After the phase alignment across the observing band, the global fringe fitting was performed \citep{1983AJ.....88..688S}, 
correcting for the residual delays and phases with respect to a chosen reference antenna. Finally, the visibility amplitudes were calibrated, taking into account corrections for atmospheric opacity computed using the measured system temperatures and gain-elevation curves of each telescope. 
The {\em RadioAstron} data were analyzed in a similar fashion as described in \cite{2016ApJ...817...96G} and \cite{2017A&A...604A.111B}. 
The calibration of the 15\,GHz data was carried out by the MOJAVE team, following the procedure described in \citet{2009AJ....138.1874L}. 

\subsubsection{VLBI imaging and model-fitting}
\label{sec:kinematics}

\begin{figure}
\centering
\includegraphics[width=\columnwidth]{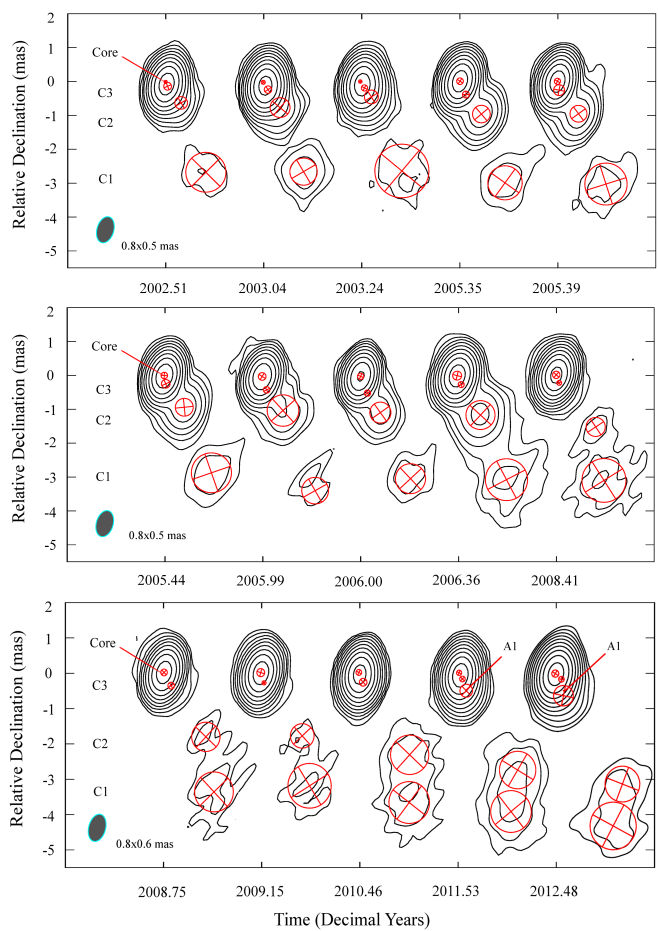}
\caption{Modeled images of blazar TXS~2013$+$370 at 15\,GHz. The 2 dimensional circular Gaussian components model the flux density distribution along the jet. The data were imaged under a uniform weighting scheme and without uv-tapering (all the visibilities weighed equally, independent of their uv-distance). The contour levels are set to 0.25,0.5, 1, 2, 4, 8, 16, 32, and 64\% Jy/beam of each image peak flux density (see Table~\ref{table:datalog}). All the images are convolved with a common beam of $0.8\times0.5$\,mas at PA $-16^\mathrm{o}$. The time corresponding to each image is indicated in the x-axis.}
\label{fig:15ghz_maps}
\end{figure}

\begin{figure}
\includegraphics[width=\columnwidth]{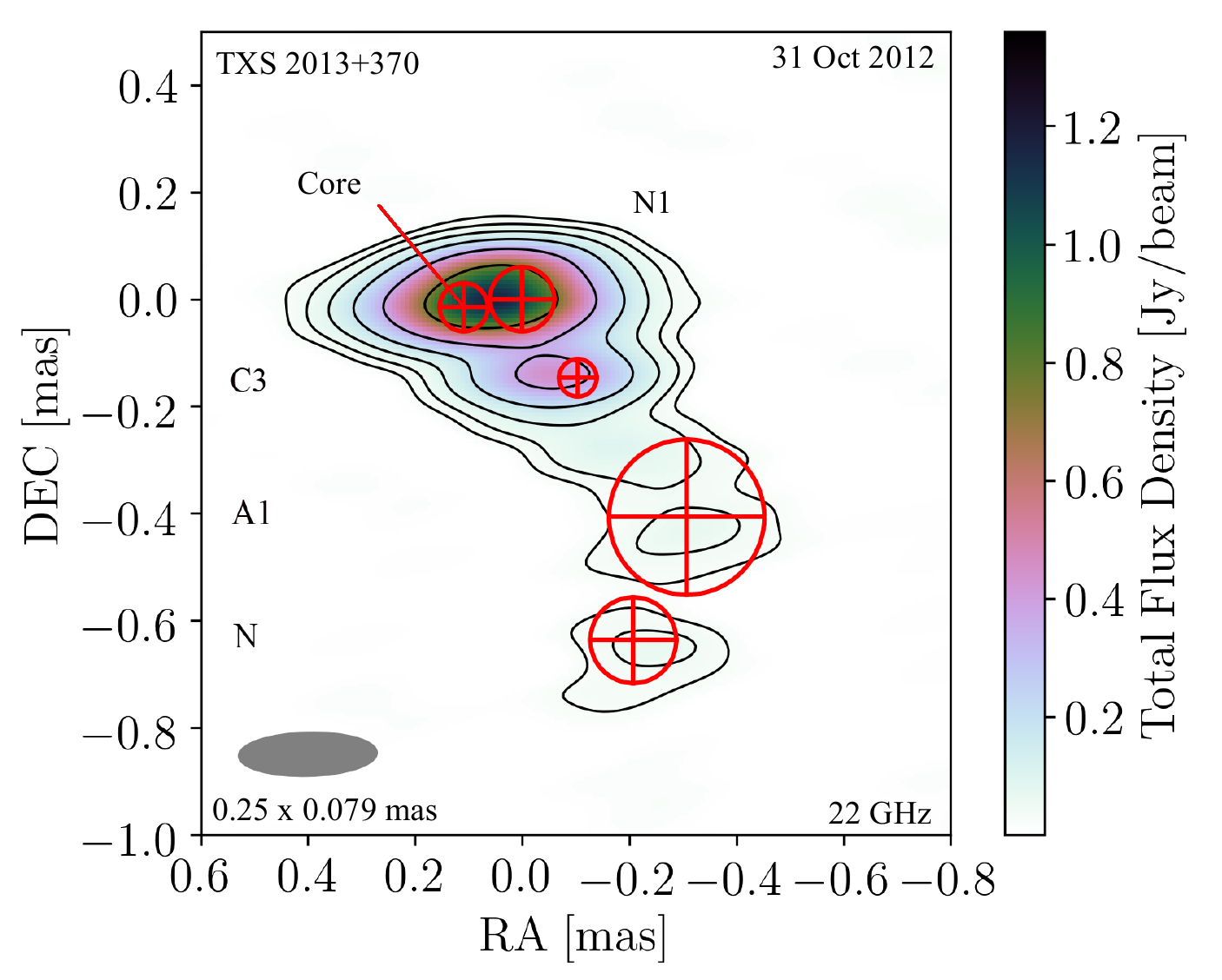}
\caption{22\,GHz model-fit and total intensity space-VLBI image of TXS~2013$+$370, obtained from a combined ground and space-VLBI array. The data were imaged under a uniform weighting scheme and without uv-tapering. The convolved beam for the displayed image is set to 0.25 $\times$ 0.08\,mas and it is oriented at a position angle of -88.6$^\mathrm{o}$. The contour levels are set to 2, 4, 8, 16, 32, and 64\% of the peak flux density of 1.22~Jy/beam.}
\label{fig:ra_maps}
\end{figure}

\begin{figure*}
\includegraphics[width=\textwidth]{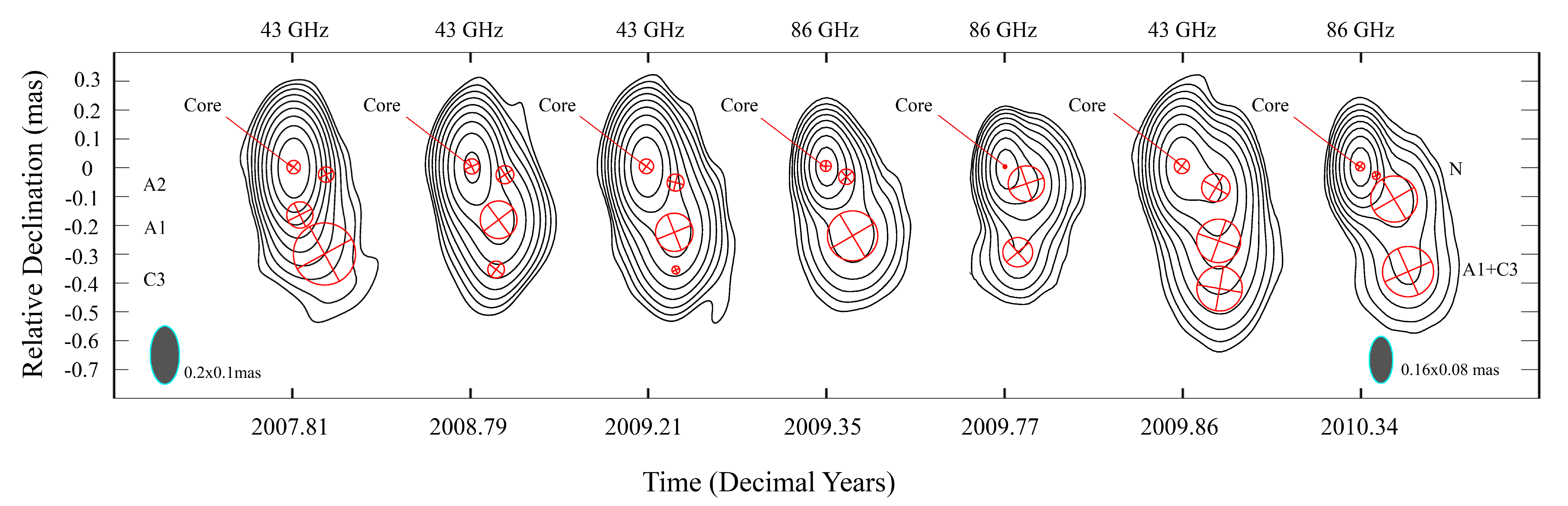}
\caption{Hybrid model fit images of TXS~2013$+$370 at 86 and 43\,GHz. The data were imaged using uniform weighting, without uv-tapering. At 43\,GHz, the images are convolved with a common restoring beam of $0.2\times0.1$\,mas, oriented at PA $0^{\circ}$, and the contour levels are set to 0.3, 0.6, 1.2, 2.4, 4.8, 9.6, 19.2, 38.4, and 76.8\% of the peak flux density (see Table~\ref{table:datalog}); at 86\,GHz, images are restored with a beam of  0.16$\times$0.08\,mas, $0^{\circ}$, and the contour levels are set to 0.5, 1, 2, 4, 8, 16, 32, and 64\,\% of the peak flux density (see Table \ref{table:datalog}). The two beams are displayed on the left and right corners respectively. The time stamp for each image is indicated on the x-axis.}
\label{fig:86-43maps}
\end{figure*}

Frequency and time-averaged data were imported to DIFMAP \citep{1994BAAS...26..987S} for the imaging and Gaussian model-fitting. Before imaging, the visibilities were carefully inspected and spurious data points were flagged. Using the {\tt CLEAN} algorithm \citep{1974A&AS...15..417H} and {\tt SELFCAL} procedures, 
which are implemented in DIFMAP, we created the radio images of the source. 
While imaging the ground-based data at 15, 43, and 86\,GHz was straightforward thanks to a large number of stations, the imaging of the {\em RadioAstron} 22\,GHz data was more challenging due to the limited uv-coverage. 
Five stations (including the space telescope) provided useful data (Table~\ref{table:datalog}). 
Robledo~70\,m recorded only right-hand circular polarization while the space telescope recorded only left-hand circular polarization at 22\,GHz (other telescopes recorded both polarizations), which resulted in no space-ground fringes to Robledo~70\,m. The amplitude calibration of the Jodrell Bank Mark\,II telescope was corrupted for an unknown technical reason. A Gaussian source model derived from the near-in-time 43\,GHz data was used to correct the amplitude of Mark\,II telescope data 
before producing the image with several iterations of {\tt CLEAN} and {\tt SELFCAL}.

For all data sets, the jet brightness distribution was then parameterized by fitting two-dimensional Gaussian components to the fully calibrated visibility data by using the {\tt MODELFIT} algorithm, which is implemented in DIFMAP. 
The uncertainty in the component positions was set to one-fifth of the beam size if the component's size was smaller than the equivalent circular beam $b=\left(b_\mathrm{max}b_\mathrm{min}\right)^{1/2}$,  otherwise, we assumed one-fifth of the component Full Width at Half Maximum (FWHM) as an estimate of the uncertainty. 
The uncertainty in the position angle, PA, was calculated based on the positional uncertainty $\Delta X$
using the trigonometric formula $\Delta PA=\arctan(\Delta X/r)$, with $r$ the radial separation in mas.
For the component flux density and FWHM, we adopt an uncertainty of 10\%, following \cite{2009AJ....138.1874L,2013AJ....146..120L} and \cite{2016A&A...586A..60K} respectively. 
The resulting images are shown in Fig.~\ref{fig:15ghz_maps},\ref{fig:ra_maps}, \ref{fig:86-43maps},  while the parameters of the Gaussian components are reported in Tables~\ref{table:kinem1}, \ref{table:kinem3}, \ref{table:kinem4}, and \ref{table:kinem5}.

Since the source is located very close to the Galactic plane, where the column density of the interstellar medium is relatively high, there is the possibility for the images to be affected by interstellar scattering.
In order to verify the impact of scattering in TXS~2013$+$370, we investigated two possible observational effects that could be introduced by it: angular broadening and fast flux density variations. By analyzing the dependence of the angular size with frequency and based on the variability properties inferred from an Effelsberg monitoring at 5\,GHz, we conclude that none of these effects play a dominant role at the frequencies considered in our study. The detailed analysis is presented in Appendix A.

\subsection{\textit{Fermi}-LAT data analysis}
\label{subs:fermi}

\begin{figure*}
\centering
\includegraphics[width=0.65\textwidth]{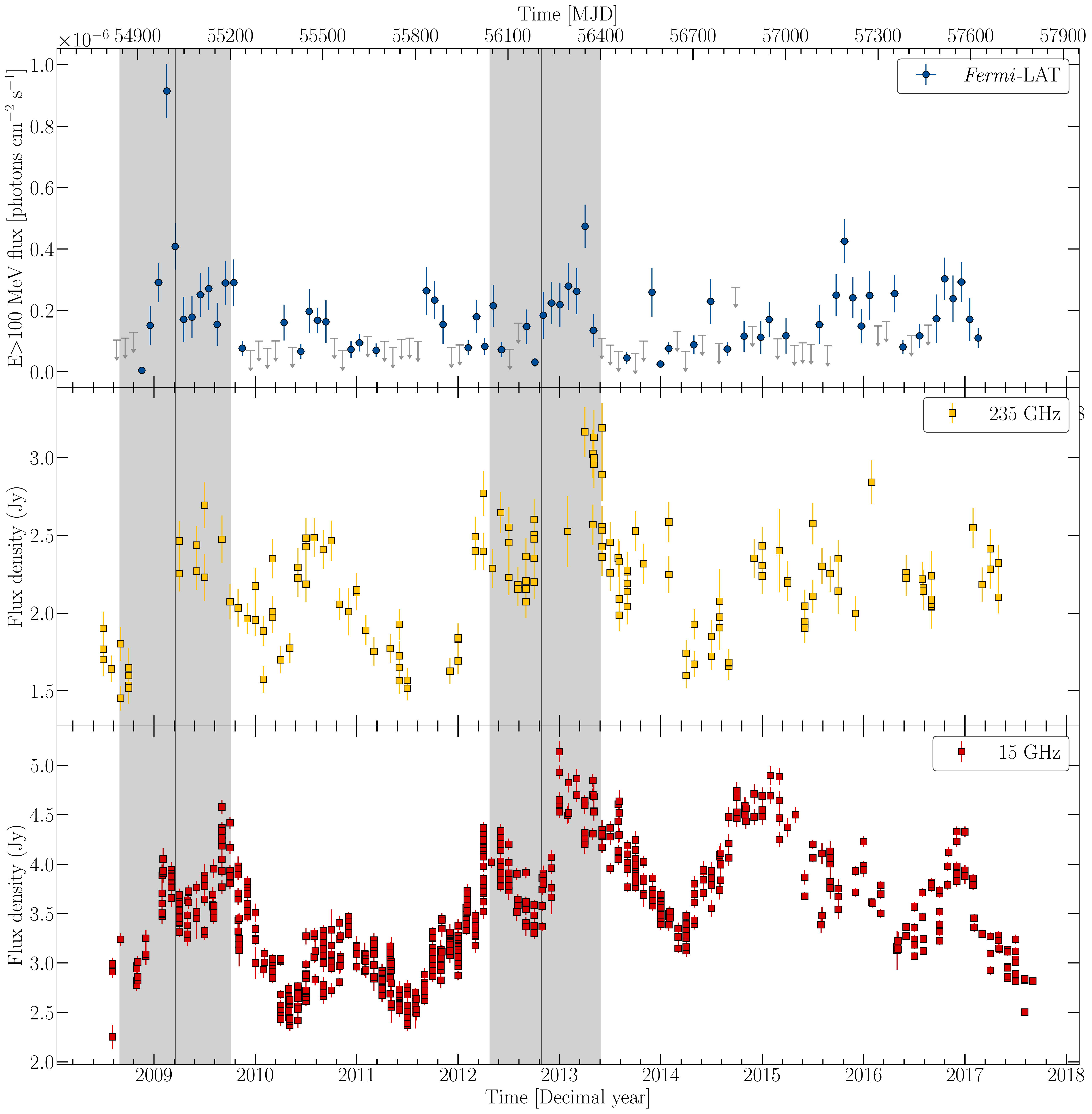}
\caption{Light curves of the blazar TXS~2013$+$370 at different frequencies. From top to bottom: \textit{Fermi}-LAT 0.1 -- 300~GeV with 30\,days binning, 235\,GHz SMA and 15\,GHz OVRO light curves, with flux plotted vs. time. The shadowed areas indicate the ejection time range of the new jet components N (see Fig.~\ref{fig:86-43maps}) and N1 (see Fig.~\ref{fig:ra_maps}) respectively. The vertical lines designate the estimated ejection time, whereas the width of the shadowed areas indicates the uncertainty of this estimation, based on the uncertainty of component A1. We adopt for the N as ejection time the peak of the 43\,GHz core flare, whereas for N1, we take the appearance time from the 22\,GHz VLBI image. We set for N and N1 the same ejection time uncertainty as the most well-defined feature C2, owing to lack of data.}
\label{fig:light-curves}
\end{figure*}

The Large Area Telescope (LAT) is a pair-conversion detector sensitive
to $\gamma$ rays from below 20\,MeV to more than 300\,GeV. It was 
launched on June 11, 2008 as the main scientific instrument on board the \textit{Fermi Gamma-ray Space Telescope} \citep{Atwood2009}. 
We used the Python package \texttt{Fermipy} \citep{2017ICRC...35..824W} throughout the analysis and assumed as a starting sky model the \textit{Fermi}-LAT third source catalog (3FGL) \citep{2015ApJS..218...23A}. 
We consider a Region of Interest (ROI) of 10$^\circ$ around the target position 
and include in the model all point sources from the 3FGL within 15$^{\circ}$ of the ROI center, together with the corresponding model for the Galactic and isotropic diffuse emission 
(\texttt{gll\_iem\_v06.fits} and \texttt{iso\_P8R2\_SOURCE\_V6\_v06.txt}, respectively). 
We performed a binned analysis with 10 bins per decade in energy and 0.1$^{\circ}$ binning in space, in the energy range 0.1-300\,GeV. We first performed a likelihood analysis over the full-time range considered here, 
i.e., 2008.58 15:43:36.000 UTC to 2017.17 15:43:36 UTC. 
We fit the ROI with the initial 3FGL model, freeing all the parameters of the target source and the normalization of all sources within 5$^{\circ}$ of the ROI center. 
Since our data set more than doubles the integration time with respect to the 3FGL catalog, 
we look for new sources with an iterative procedure. We produce a map of Test Statistic (TS). 
The TS is defined as $2\log(L/L_0)$ where $L$ is the likelihood of the model with a point source at the target position, 
and $L_0$ is the likelihood without the source. A value of TS=25 corresponds to a significance level of 4.2$\sigma$~\citep{Mattox1996}. 
A TS map is produced by inserting a test source at each map pixel and evaluating its significance over the current model. 
We look for TS$>$25 peaks in the TS map, with a minimum separation of 0.3$^{\circ}$ and add a new point source to the model for each peak, 
assuming a power-law spectrum. We then fit the ROI again and produce a new TS map. 
This process is iterated until all significant excesses are modeled out~\footnote{In our case, the iterative source finding procedure resulted in the addition of 19 new point sources in total. 
This relatively high number is justified by the increased integration time of our analysis with respect to the 3FGL, and by the fact that the source lies close to the Galactic plane. 
The latter implies that some of the point sources added might correspond to un-modeled diffuse emission. However, none of these sources has a flux comparable to the target source.}. 
We also perform a localization analysis on the target source and all new sources with TS$>$25 found in the ROI.

For the variability analysis, we perform a likelihood fit in each time bin, using the average model as a starting point. 
We first attempt a fit leaving the full spectrum of the target source, which is described by a LogParabola, free to vary. 
If the statistics do not allow the fit to converge or results in a non-detection (TS$<$25), we fix all parameters except the target source's normalization. 
We consider the target source to be detected if TS$>10$ in the corresponding bin and the signal-to-noise ratio (i.e., flux over its error) 
in that bin is larger than two. If this is not the case, we report a 95\% confidence upper limit.

\subsection{Single-dish radio light curves}

Single-dish radio data, contemporaneous to the VLBI observations, were provided by the 40-m telescope of the Owens Valley Radio Observatory (OVRO) at 15\,GHz \citep{2011ApJS..194...29R}.
The data at 235\,GHz were obtained with the 8-element Submillimeter Array (SMA) \citep{2007ASPC..375..234G} for the period 2008--2017. 

As shown in Fig.~\ref{fig:light-curves}, where the OVRO, SMA, and \textit{Fermi} light curves are presented, the source was quite active during our monitoring period, and several flaring episodes occurred both at low and high energies. The investigation of possible correlated variability between the considered energy bands is presented in Sect.~\ref{subs:multi_band_var}.

\section{Data analysis and results}
\label{sec:analysis_results}

\subsection{Source structure and jet kinematics}

Our imaging at 15, 22, 43, and 86\,GHz shows that TXS~2013$+$370 features a bright core and a bent jet extending to 5\,mas at the lowest frequency. We determined the kinematics of the individual 
jet features using the parameters of the Gaussian components 
derived from the model fits of each data set.
The component cross-identification between the observing epochs (and frequencies) was done by comparing their positions, flux densities, and sizes. 
The angular proper motion $\mu$ of the components was computed through linear fits of their radial core separation as a function of time. The apparent speed $\beta_\mathrm{app}$ is related to $\mu$ and the intrinsic speed $\beta$ as \citep{1995PASP..107..803U, 1966Natur.211..468R}:

\begin{equation}
\beta_\mathrm{app}=\frac{\beta \sin \theta}{ 1-\beta \cos \theta} = \frac{\mu D_L}{c(1+z)},
\label{eq:beta_app}
\end{equation}

\noindent where $\theta$ is the viewing angle of the jet, $\mu$ is the proper motion in rad~s$^{-1}$, $D_{L}$ is the luminosity distance in m, $z$ the source redshift and $c$ is the speed of light in m\,s$^{-1}$ \citep[e.g.,][]{1999astro.ph..5116H}.

\paragraph{15 GHz:}
Figure~\ref{fig:15ghz_maps} presents the 15\,GHz contour images with super-imposed circular Gaussian model-fit components, while the parameters of each component are listed in Table~\ref{table:kinem1}.

\begin{figure*}
\centering
\includegraphics[scale=0.2]{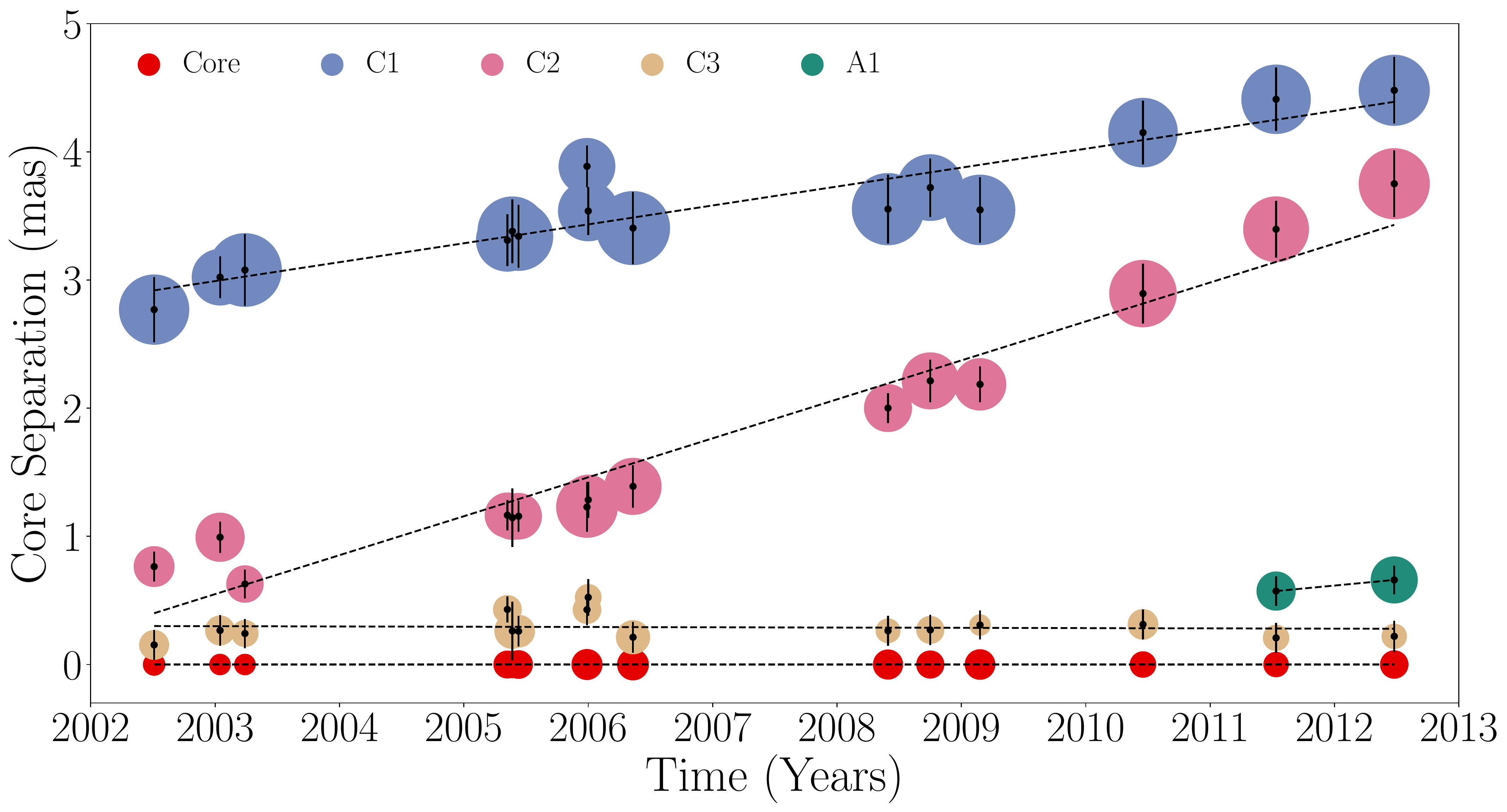}
\includegraphics[scale=0.2]{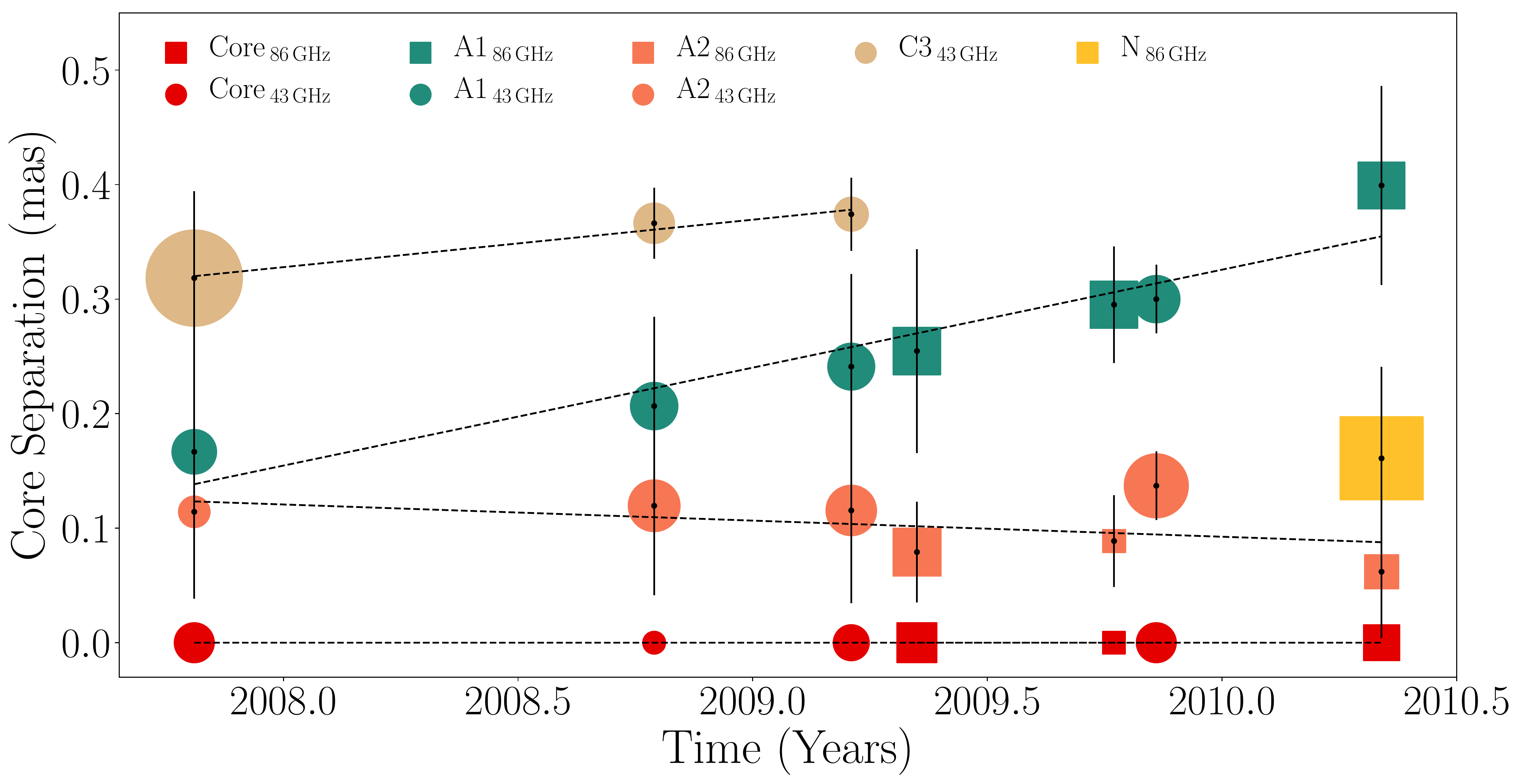}
\includegraphics[scale=0.2]{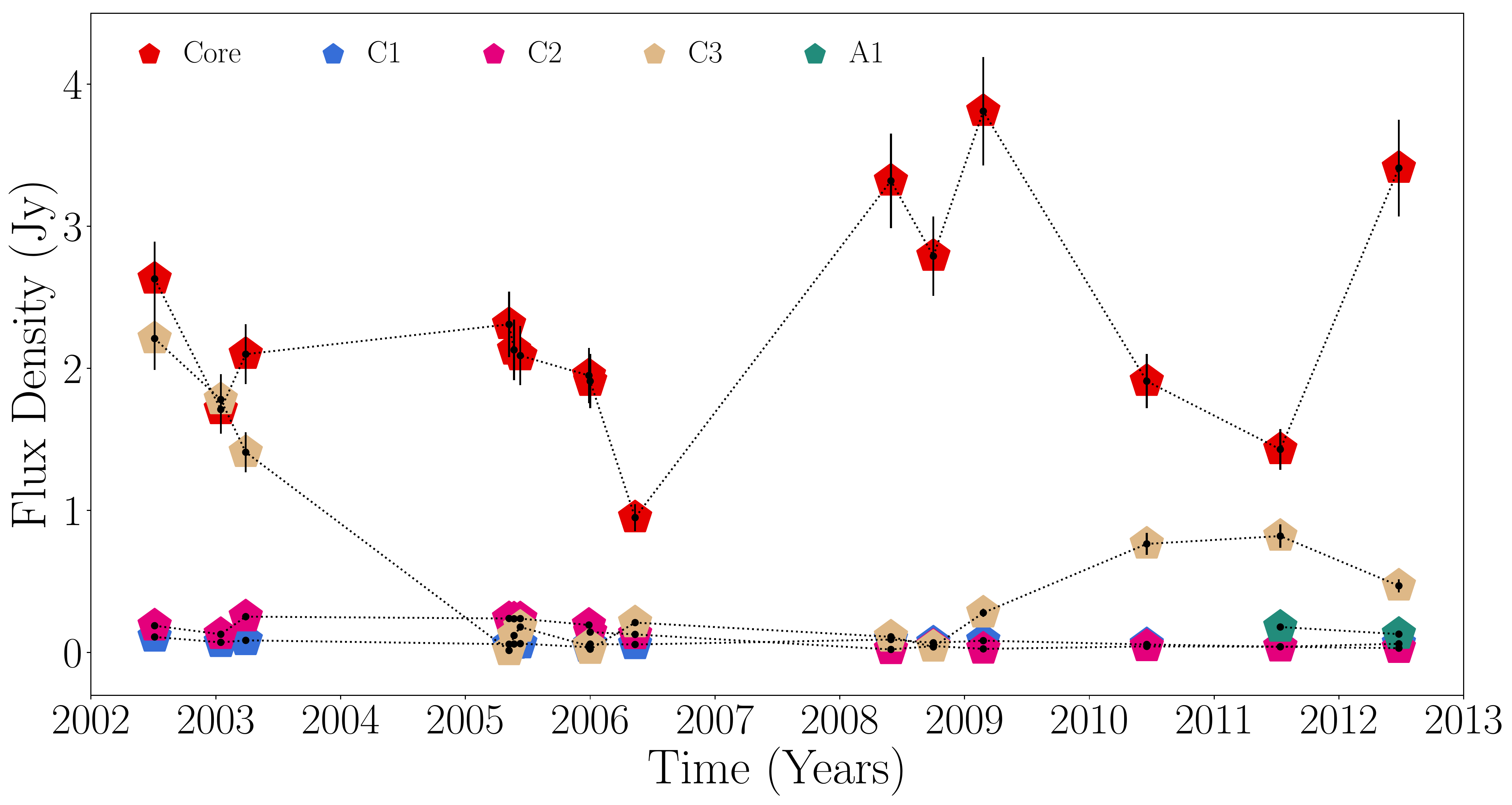}
\includegraphics[scale=0.2]{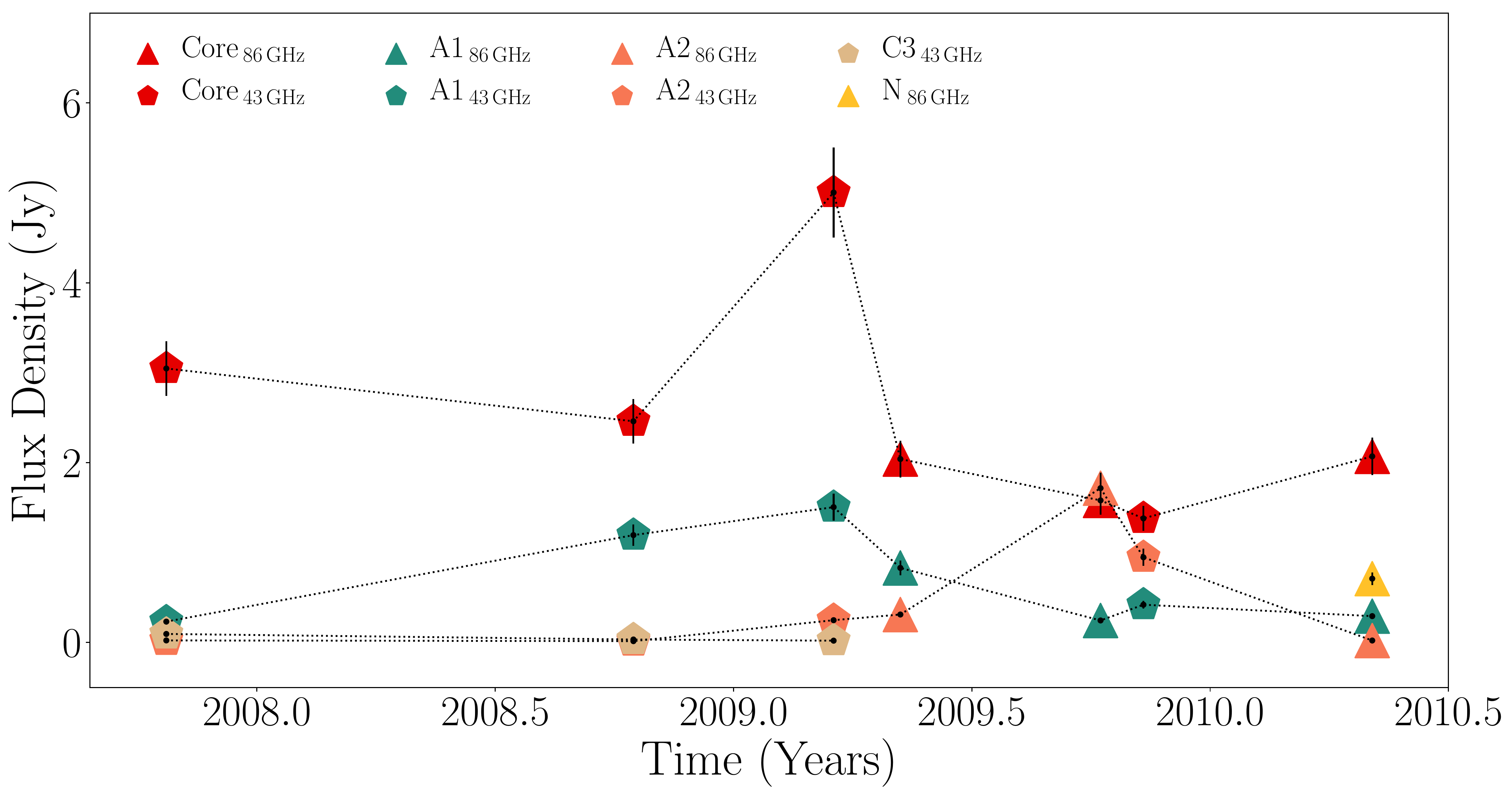}
\caption{Time evolution of the core distance (top panels) and flux density (bottom panels) of model-fit components at 15\,GHz (left panels) and at 43 and 86\,GHz (right panels). In the top panels, the size of the data points is set to ${\sim}$1/2 of the Gaussian components size. Note here that the time range of 15 and 43-86\,GHz data is different. The flux density of the core component showed a maximum in early 2009 at all frequencies. Based on the observational data, in this time range a new component emerges from the core. This new knot seems to "light up" component C3 on its passage, increasing the flux density of the latter.
}
\label{fig:kinematics}
\end{figure*}

At this frequency, the source is well modeled by four circular Gaussian components, a core component (labeled as Core) and three jet features (C3, C2, C1), numbered in order of decreasing distance from the core. During the observing interval 2002-2012, component C3 appears quasi-stationary \citep{1985ApJ...295..358L}, oscillating around an average distance of $r \sim 0.2$\,mas from the center (Fig.~\ref{fig:kinematics}, top-left panel). 
Numerous studies have shown that when a moving shock passes through a stationary one, the latter could be displaced in position for a short time and then return to its initial position \citep[see][and references therein]{2015A&A...578A.123R}. Component C3 seems to exhibit such behavior. 
The features C2 and C1 are, instead, moving components,
separating from the core with apparent superluminal speed (see Table~\ref{table:kinematics}).
From our kinematic analysis, we infer that C2 is the fastest component of the jet, 
moving with an apparent speed of $\beta_\mathrm{app}=13.8 \pm 0.9$. 
We note that a quadratic fit was performed by \cite{2016AJ....152...12L} to describe the motion of C2. Their speed ($\beta_\mathrm{app}=14.51\pm0.24$)
is consistent with our result within the measurement uncertainty.
Component C1 is the second fastest feature of the jet, moving at a speed of
$\beta_\mathrm{app}=7.0 \pm 0.8$. 
We also note that a new component, labeled A1, becomes visible at 15\,GHz in the last two VLBI epochs (after 2011.53), downstream of C3. See the last two images in Fig.~\ref{fig:15ghz_maps}, and the component 
separation as a function of time in Fig.~\ref{fig:kinematics} (top left panel). The observation of A1 in only two epochs limits the accuracy of its speed determination. Based on the 15\,GHz data points, we obtain an apparent speed of $\beta_\mathrm{app}=4.2 \pm 11.7$, which is in agreement with the value derived for the same component at 43 and 86\,GHz (see below).

\paragraph{22 GHz:}
Figure~\ref{fig:ra_maps} shows the 22\,GHz image of the inner jet region (pc-scale) of TXS~2013$+$370 as observed by space-VLBI with {\em RadioAstron} \citep[details in][]{2014cosp...40E3161S}. The image shows a core, which is elongated in the East-West direction and along the major axis of the elliptical observing beam. The jet appears to be propagating towards a South-West direction, with a slight curvature towards the southern direction beyond $r=0.4$\,mas. The overall morphology of the VLBI structure is in good agreement with the 43 and 86\,GHz images (Fig.~\ref{fig:86-43maps}). Two components can model the central region, the core and a new feature, labeled as N1 (the second grey area in Fig.~\ref{fig:ra_maps}). The appearance and the propagation of N1 probably is connected with the close-in-time increased activity seen in the radio and $\gamma$-ray bands (see Fig.~\ref{fig:light-curves}).
Further downstream, three additional Gaussian components represent the jet. Based on their core separation and flux density, we identify these features with the components C3 and A1 at 15\,GHz, and with component N at 86\,GHz (see below).

\paragraph{43 \& 86 GHz:}
Figure~\ref{fig:86-43maps} shows the model-fit images of the source at high radio frequencies. Only the most compact regions in the inner jet are visible. The source structure can be modeled by a compact core and three jet components, the innermost of which, A2, is not visible at lower frequencies. The core separation of each component as a function of time is shown in 
Fig.~\ref{fig:kinematics} (top-right panel), while the component parameters are reported in Tables \ref{table:kinem4} and \ref{table:kinem5}. Component A2, located at $r \sim 0.1$\,mas from the core, appears to be quasi-stationary, showing some backward motion after 2009, probably due to the passage of a newly ejected feature, N, which becomes well separated from A2 in the final epoch at 86\,GHz. The appearance of N followed a strong increase (by a factor of ${\sim}2$) of the core flux density, observed in March 2009 at 43\,GHz (Fig.~\ref{fig:kinematics}, bottom-right panel). At around this time, a prominent $\gamma$-ray flare was also observed by \textit{Fermi}-LAT (see Fig.~\ref{fig:light-curves}).
Further downstream, component A1, which is resolved at 15\,GHz only after 2011, moves with an apparent speed $\beta_\mathrm{app} = 4.0 \pm 0.7$. The motion of the outermost component, C3, is not well constrained at 86\,GHz, since in the region occupied by C3 the jet becomes faint and partially resolved, and after 2009 there is a possible blending between C3 and A1. However, the 43\,GHz data show a well defined C3 component in the first three epochs, from which we infer a proper motion $\mu=(0.07\pm 0.02)$\,mas/yr, which corresponds to an apparent speed of $\beta_\mathrm{app} = 3.3 \pm 1.1$. This result is comparable to the speed of the nearest upstream component A1. As discussed above, C3 appears quasi-stationary in the long 15\,GHz monitoring, therefore the small displacement observed at 43\,GHz in the short time is likely associated with flowing plasma crossing C3 at that time.

In summary, the source TXS~2013$+$370 shows a bent jet, curving from an east-western to a southern orientation. The VLBI core is much more variable in flux density than the jet components. These move at apparent superluminal speeds of $\sim$(3-14)\,c, indicating highly relativistic motion of the jet plasma.  Near the core, stationary components, as well as newly ejected features, are observed. The ejection times for several jet components could be estimated and are reported in Table~\ref{table:kinematics}. 
A discussion of possible reasons for the  wide range of apparent speeds and on the implications for the jet intrinsic parameters (e.g., Lorentz factor and viewing angle) is  reported in Sect.~\ref{subs:intrinsic_param}.

\begin{table}[!ht]
\caption{Kinematic parameters of all identified components at 15, 43, and 86\,GHz. Columns from left to right: (1) Component ID, (2) Observing frequency, (3) Proper motion, (4) Apparent speed (5) Ejection time} 
\label{table:kinematics}   
\centering                         
\begin{tabular}{@{}c c c c c c c c c c@{}}    
\hline\hline                
Knot & Freq.  &$\mu$ & $\beta_\mathrm{app} $ & $t_\mathrm{ej}$  \\ & (GHz) & (mas/year) & ($c$) & (year)  \\
\hline                      
A1 & 15 & 0.09 $\pm$ 0.20 & 4.2 $\pm$ 11.7 & 2005.27 $ \pm $ 1.20 \\
C3 & 15 &  - & - & quasi-stationary \\ 
C2 & 15 &  0.30 $ \pm $ 0.02  & 13.9 $\pm$ 0.9 & 2001.08 $ \pm $ 0.55  \\
C1 & 15 & 0.15 $ \pm $ 0.02 & 7.0 $\pm$ 0.8 &  1982.72 $ \pm $ 0.49   \\     
A2 & 43/86 & - & - &  quasi-stationary \\   
A1 & 43/86 & 0.09 $\pm$ 0.01 & 4.2 $\pm$ 0.5 & 2006.14 $\pm$ 0.22   \\ 
C3 & 43/86 & 0.07 $\pm$ 0.02 & 3.3 $\pm$ 1.1 & 1997--2003  \\
\hline 
\hline
\end{tabular}
\end{table}

\subsection{Location of the jet apex}
\label{subsub:apex_loc}

In this section, we aim to constrain the location of the jet apex with respect to the VLBI core. Such an estimate can be obtained through an analysis of the frequency-dependent shift of the VLBI core position, caused, in the simplest scenario, by synchrotron opacity and self-absorption. In the case of TXS~2013$+$370, however, this approach cannot be used directly, since a formal 2-D cross-correlation using close-in-time pairs of images at 15\,GHz, 43\,GHz, and 86\,GHz, yielded no measurable shifts. As we will discuss in Sec.~\ref{sec:discussion} based on the variability time lags, the core shift is indeed much smaller than our resolution limits in the considered frequency regime. 

In order to obtain an estimate of the jet apex location with respect to the VLBI core, we then follow a geometrical approach based on the investigation of the transverse jet expansion profile. At each frequency, we convolved all images with the average equivalent circular beam (0.08\,mas at 86\,GHz, 0.16\,mas at 43\,GHz, and 0.75\,mas at 15\,GHz) and created stacked images. We have not considered the 22\,GHz \textit{RadioAstron} data since there is only one epoch available. We then measured the jet width as a function of core separation in stacked images. This was done by slicing the jet pixel-by-pixel in the direction perpendicular to the jet axis, and by fitting single Gaussian profiles to the transverse intensity distribution to infer the jet width. For the error of the jet width,  we used one-tenth of the convolved FWHM.

\begin{figure}
    \centering
    \includegraphics[width=0.5\textwidth]{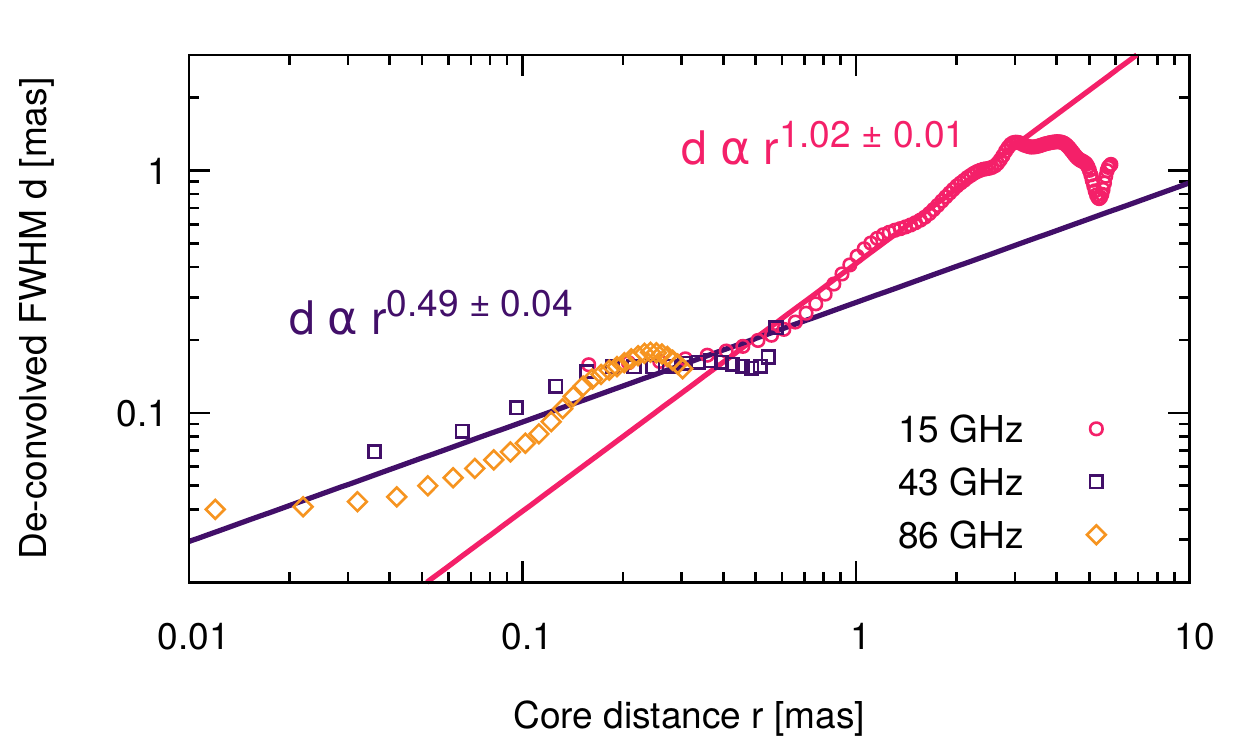}\\
        \includegraphics[width=0.5\textwidth]{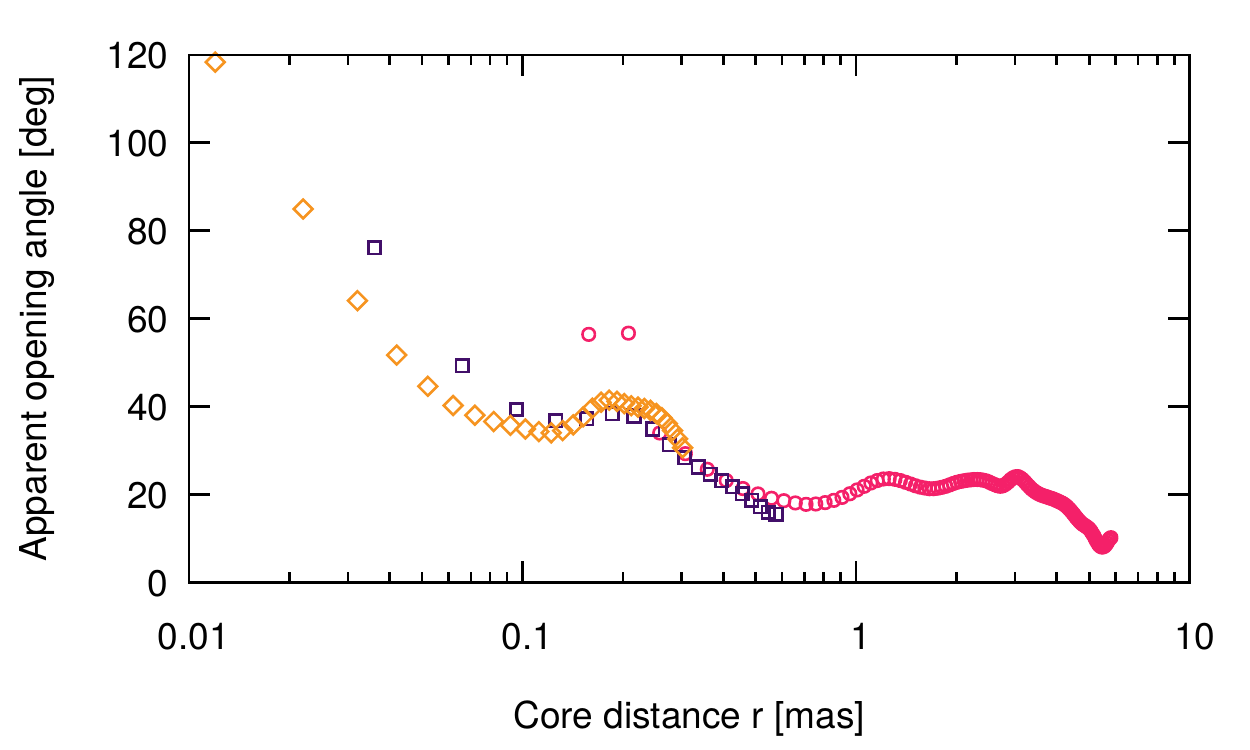}
    \caption{Top - Transverse expansion profile of the jet in TXS~2013$+$370, based on 86\,GHz, 43\,GHz, and 15\,GHz data. Error bars are assumed equal to one-tenth of the convolved FWHM and are not shown in the plot for clarity. The jet has a parabolic shape in the inner regions and a conical shape on larger scales. The transition occurs at a projected distance of $\sim$0.5\,mas from the central engine. Bottom - Apparent opening angle as a function of distance from the core.}
    \label{fig:my_label}
\end{figure}

The expansion profile is presented in Fig.~\ref{fig:my_label}, top panel. Distances along the x-axis are relative to the position of the VLBI core, which we assume as being fixed, due to the negligible core-shift. The de-convolved FWHM values inferred for the inner jet region based on the 86 and 43\,GHz data smoothly connect to those inferred at 15\,GHz for the outer regions. By fitting a power-law of the form $d=ar^b$ to the 15\,GHz data at distances between ${\sim}0.4$ and ${\sim}3$\,$\rm mas$ from the core, we infer that the jet has a conical shape ($d\propto r^{(1.02\pm0.01)}$). However, this power law does not describe well the higher frequency data, as in the inner jet we observe a flattening of the expansion profile. By fitting a power law of the same form to the 43\,GHz and 86\,GHz data, we obtain $d\propto r^{(0.49\pm0.04)}$, i.e., the jet has a parabolic shape in the proximity of the black hole. This is expected based on theoretical models for jet formation \citep[e.g.,][and references therein]{2001Sci...291...84M}, which predict the jet to be actively collimating and accelerating at its onset. While we have not considered in the fit the 15\,GHz data relative to the inner jet, the lower resolution data points also lie on the same profile. The fact that the jet is collimating on the scales probed by mm-VLBI is also evident by examining the evolution of the apparent opening angle with distance (Fig.~\ref{fig:my_label}, bottom panel). The angle decreases in the inner ${\sim}0.5$\,$\rm mas$, and then reaches a roughly constant value of ${\sim}$23$^{\circ}$ on the scales probed at 15\,GHz, until the recollimation region at a distance of $\sim4$\,$\rm mas$. As observed in other more nearby jets like M\,87 \citep{2012ApJ...745L..28A} or Cygnus\,A \citep{2016A&A...585A..33B}, for which higher spatial resolution is achieved, a single parabolic profile describes the expansion of the jet from its onset up to the parsec scales. Therefore, we take the inferred expansion law to back-extrapolate the location of the jet apex, which should have an approximately zero width, with respect to the 86\,GHz core. Such a method has been applied in literature before \citep{2011ApJ...735L..10A, 2016A&A...586A..60K}, considering the core size as a reference and then assuming a conical expansion between the black hole and the core. Similarly, we can compute the distance $R$ it takes for the jet to reach the width of the core, assuming the inferred parabolic expansion rate. Based on the analysis of the stacked image at 86\,GHz, the jet width at the location of the emission peak is $d_{c} \leq(0.04\pm0.01)$\,$\rm mas$, from which we infer that the jet apex is located at a distance of  $R\leq  ((0.019\pm0.009)/\sin{\theta})$\,$\rm mas$, or  $((0.146\pm0.07)/\sin{\theta})$\,$\rm pc$ upstream, where $\theta$ is the viewing angle of the jet.

\subsection{Multi-band variability}
\label{subs:multi_band_var}

\begin{figure*}[!t]
\minipage{0.32\textwidth}%
  \includegraphics[width=\linewidth]{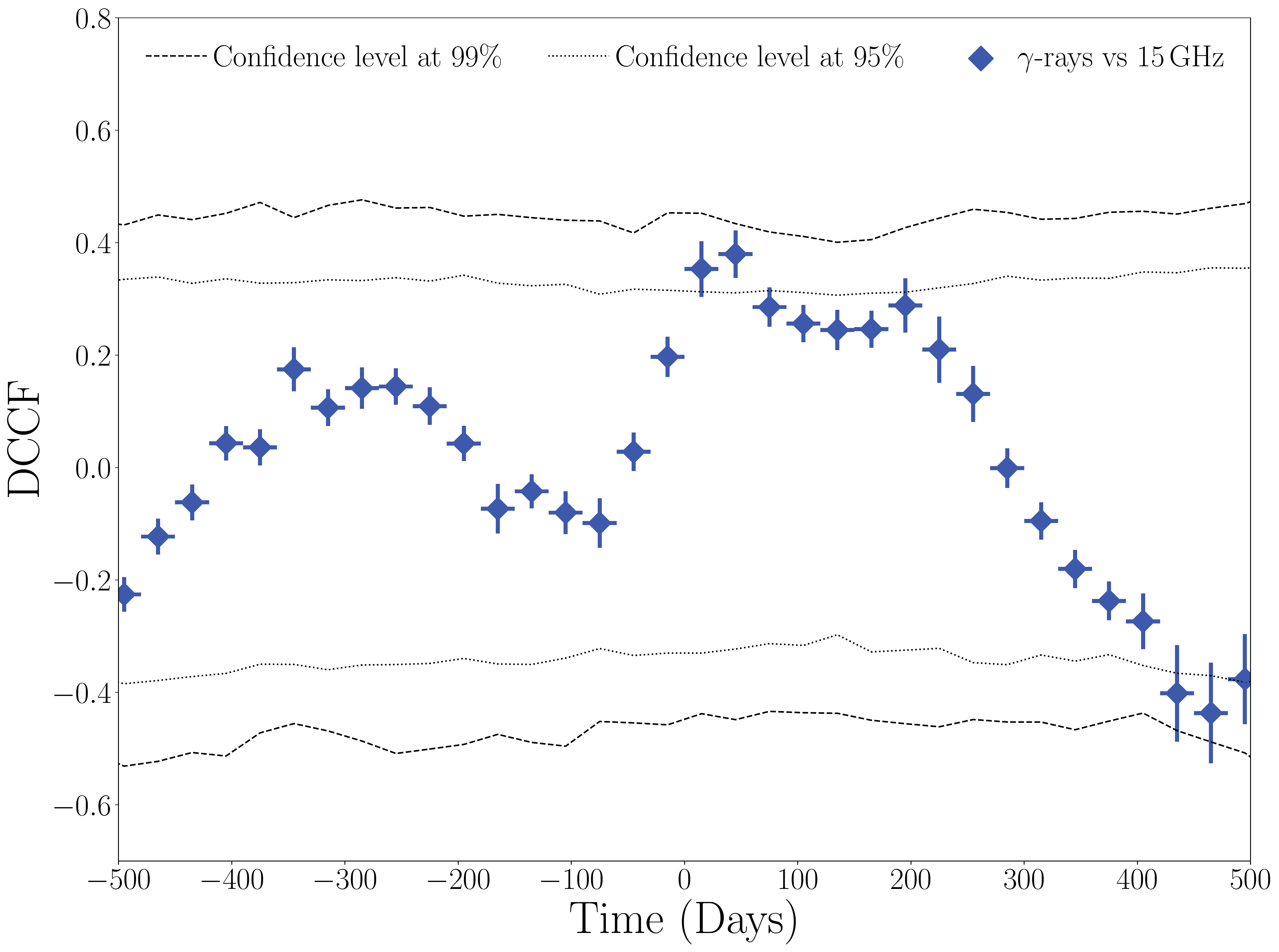}
\endminipage
\hspace{0.5cm}
\minipage{0.32\textwidth}
  \includegraphics[width=\linewidth]{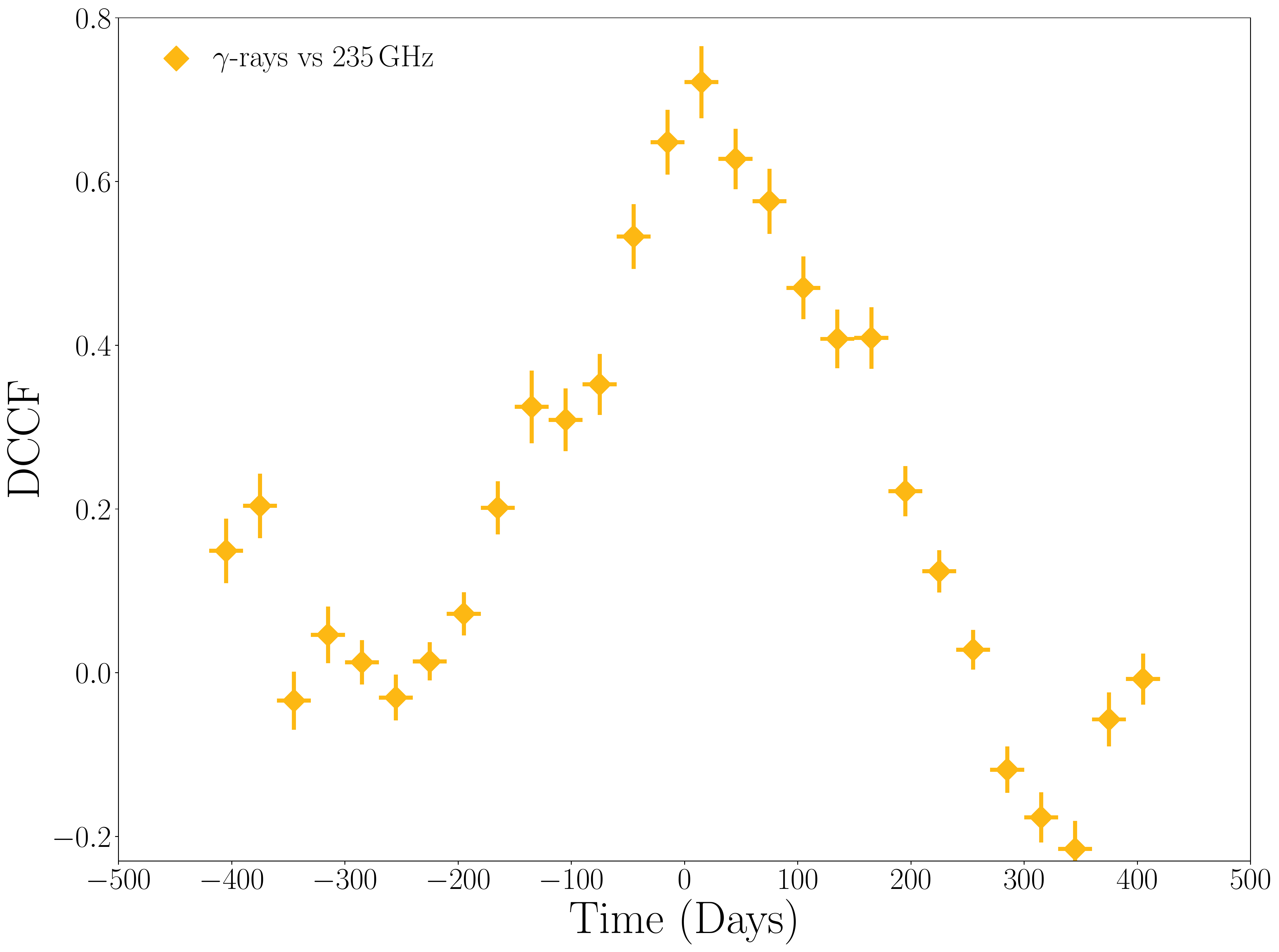}
\endminipage\hfill
\minipage{0.32\textwidth}
  \includegraphics[width=\linewidth]{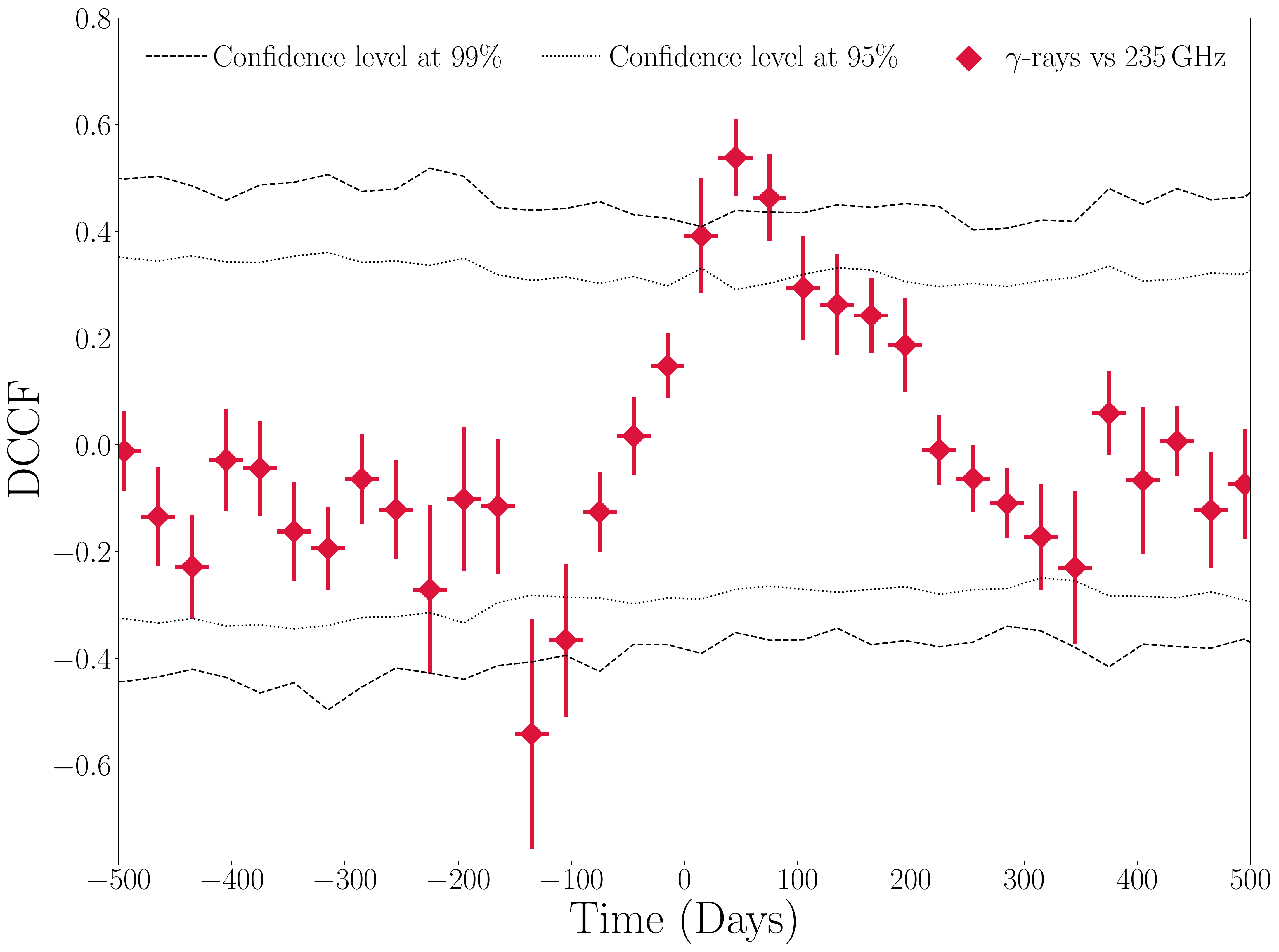}
\endminipage\hfill
\caption{DCCF results between the $\gamma$-ray and 15\,GHz light curves (left), 15-235\,GHz (middle) and $\gamma$-rays-235\,GHz (right). Positive time lags indicate that $\gamma$-ray activity leads the activity in radio. The significance of the correlations is displayed by a dotted line for the $2\sigma$ and by the dashed line for the $3\sigma$ level.}
\label{fig:gamma-235}
\end{figure*}

In order to investigate possible correlated variability across the observing bands, we applied a discrete cross-correlation function \citep[DCCF;][]{1988ApJ...333..646E}.
To test the significance of the DCF peak, we simulated the light curves using the \cite{2013MNRAS.433..907E} method implemented by \cite{2015arXiv150306676C}. We first determined the power spectral density (PSD) slope of the observed $\gamma$-ray light curve. Since we ignored the upper limits in the observed $\gamma$-ray light curve, the data are not evenly sampled. We therefore interpolated them using the cubic spline method in R\footnote{\url{https://stat.ethz.ch/R-manual/R-devel/library/stats/html/splinefun.html}}. 
As the data are manipulated via interpolation, we tested our PSD results using simulations. Specifically, we first calculated the slope and normalization of the original  PSD to determine its slope and normalization. The PSD slope and normalization were then used as variables in the simulation algorithm. The next step sampled the simulated light curves at the same bin width as the observed ones. Finally, for each pair of slope and normalization, we simulated a hundred light curves. The PSD slope of each simulated light curve was then averaged to determine its mean and standard deviation. The detailed step-by-step procedure is described by \cite{2016A&A...590A..61C}.
Our analysis suggests that the observed source variability can be well described by a PSD slope of $-(0.9\pm0.2)$. The method also takes into account the underlying probability density function (PDF) of the given light curve. We simulated a total of 5000 light curves using the best-fit PDF and PSD parameters. 
The simulated data have been correlated with the observed radio data at 235\,GHz, and 15\,GHz bands to calculate the 99$\%$ confidence levels.
A correlation is found between the $\gamma$-ray and the 235\,GHz light curves, with the radio following the high-energy activity by $\left(45 \pm 30 \right)$\,days, whereas between 235 and 15\,GHz the time lag is $\left( 15 \pm 15\right)$\,days (see Fig.~\ref{fig:gamma-235}). The confidence level of this correlation exceeds 99$\%$. The correlation between 15\,GHz and $\gamma$ rays exceeds the 95\% significance level with a delay of  $\left(45 \pm 30 \right)$\,days, which is above the 2$\sigma$ statistical threshold. In order to cross-check the degree of correlation, we applied a Spearman's correlation test \citep{10.2307/1412159}. The Spearman’s rho correlation coefficient ($\rho$) is a statistical measure of the strength of a monotonic relationship between paired data, which in this case corresponds to the differences between the fluxes of the light curves. Specifically, we used two light curves at a time, assuming as a reference the delayed one, as established by the DCCF analysis. Thus, we shifted the other data set from 0 to 200\,days with a step of 1 day and taking into account a coherence time of 2\,days. For each iteration, we calculated the P-value to quantify the significance of this statistical hypothesis. We performed this procedure for the pairs $\gamma$ ray-235\,GHz, $\gamma$ ray-15\,GHz and 15\,GHz-235\,GHz. Errors in the time-lag estimation have been set equal to the data sampling interval. The time lag between $\gamma$-rays-15\,GHz was calculated based on the Spearman's Rho results of the other two data sets. The obtained time lag between $\gamma$ ray-15\,GHz is $\left( 56 \pm 30 \right)$\,days with P-value=$1.5\times 10^{-3}$, for $\gamma$ ray-235\,GHz is $\left( 52\pm 30 \right)$\,days with P-value=$2.1\times 10^{-4}$, whereas the time lag between 235 and 15\,GHz was found to be $\left( 11\pm 6 \right)$\,days with P-value=$9.5\times 10^{-9}$ . As a sanity check, we searched for correlated activity between the 15\,GHz data set and itself. The result was, as expected, $\rho$=0.997 with P-value=0 at 0\,days shift. 
Both methods, the DCCF and the Spearman’s rho test, showed that the high-energy emission leads the activity in the radio band (see a summary of the results in Table 4).  This indicates that the high-energy event may have occurred in a region that is opaque to radio waves \citep{2014MNRAS.441.1899F, 2010ApJ...722L...7P} or, alternatively, that a denser source of seed photons for inverse Compton scattering is present upstream of the main millimeter-wave emission region.

In summary, we conclude that the $\gamma$ rays lead the variability. In the following analysis, we will assume the mean time lag for each pair of light curves. Hence, we consider that the $\gamma$-ray variability is ahead of 235\,GHz by $\left(49\pm 30 \right)$\,days. The 15\,GHz data lag behind by another $\left( 13\pm11 \right)$\,days (relative to the 235\,GHz data), while the lag between 15\,GHz and $\gamma$ rays is $\left(51\pm30 \right)$\,days (see column 4 in Table \ref{table:gamma_loc}).

\begin{table*}[!t]
\caption{DCCF, Spearman’s rho test results and linear distance between $\gamma$-ray and VLBI core. Columns from left to right: (1) Data set pair, (2) Time lag estimation from DCCF, (3) Significance level of the DCCF results, (4) Time lag estimation from the Spearman's rho test, (5) P-Value, (6) Mean time lag}    
\label{table:gamma_loc}
\centering 
\begin{tabular}{@{}cccccc@{}} 
\hline\hline
 & DCCF & Significance Level & Spearman's rho & P-Value & Mean     \\
Data set & Time lag (days) & \% & Time lag (days) & &  Time lag (days) \\ 
\hline
$\gamma-15$ & 45 $\pm$ 30 & 95  & 56 $\pm$ 30 & 1.5 $\times 10^{-3}$ & 51 $\pm$ 30   \\
$\gamma-235$ &  45 $\pm$ 30 & 99 & 52 $\pm$ 30 & 2.1 $\times 10^{-4}$ & 49 $\pm$ 30   \\
$235-15$ & 15 $\pm$ 15  &  99 &  11 $\pm$ 6 & 9.5 $\times 10^{-9}$ & 13 $\pm$ 11 \\
\hline
\end{tabular}
\end{table*}

\section{Discussion}
\label{sec:discussion}

\subsection{Intrinsic jet parameters}
\label{subs:intrinsic_param}

The diversity of apparent speeds found in TXS~2013$+$370, with values ranging from moderate ($\beta_\mathrm{app}\sim 3$) to high ($\beta_\mathrm{app}\sim 14$, see Table~\ref{table:kinematics}), is quite common among extragalactic jets \citep[e.g.,][]{2013A&A...551A..32F,2015A&A...578A.123R,2016A&A...586A..60K,2017ApJ...846...98J} and could result from several effects. The VLBI images show a jet bending from west to south at a distance of 0.1-0.2\,mas from the core, and then again towards west on the scales probed at 15 GHz. Therefore a first hypothesis is that the apparent speed variations are geometry-dependent, with the apparent speed increasing when the jet points closer to the line of sight. By considering the position angles of the moving components in Tables~\ref{table:kinem1}, \ref{table:kinem4}, \ref{table:kinem5}, there is indeed an indication that the slowest features are those moving towards south (A1, C3), while the highest speed is observed for C2, which follows a trajectory towards west and south-west. 
Another possibility is that the Lorentz factor is not constant along the jet, but increases as a function of distance. This hypothesis is supported by the observation of active jet collimation in the inner 0.5\,mas of the jet (see Sect.~\ref{subsub:apex_loc}), indicating that the terminal Lorentz factor is not yet reached on the scales probed by mm-VLBI, where the slowest apparent speeds are measured. 
Ultimately, it is likely that both geometrical effects and intrinsic variations of the bulk speed cause the observation of this wide range of apparent speeds in TXS~2013$+$370.

Based on this premise and the results of the kinematic analysis, in the following we estimate some of the intrinsic jet parameters. 
The observation of a maximum speed of ${\sim}$13.9\,c for C2 implies that at a distance of ${\sim}$1-2\,mas from the core the flow has a minimum Lorentz factor $\Gamma_{\mathrm{min}}=14.0 \pm 0.8$, since $\Gamma_{\rm min}$ is expressed as

\begin{equation}
    \Gamma_\mathrm{min}= \left( \beta^{2}_\mathrm{app} +1 \right)^{1/2}.
    \label{eq:gamma_min}
\end{equation}

The viewing angle that maximizes the apparent speed for a given Lorentz-factor is called the critical viewing angle $\theta_{\rm c}$:

\begin{equation}
    \theta_\mathrm{c} = \sin^{-1} \left( 1/\Gamma_\mathrm{min}\right).
\end{equation}

For C2 we obtain $\theta_{\rm c}=(4.1 \pm 0.2)^{\circ}$. The critical viewing angle of the fastest moving component is often assumed in the literature to be equal to the characteristic jet viewing angle \citep[e.g.,][]{1994ApJ...430..467V,1997ApJ...476..572L,Cohen_2007}. This angle could vary along the jet and be larger for those regions of the jet which are pointing towards south, and for which the slower speeds are measured (e.g., $\theta_c \simeq 16^{\circ}$ for components A1 \& C3). However, for this paper we are mostly interested in the viewing angle of the inner jet and of the core region, where the plasma moves in a direction similar to C2. Therefore we will adopt this viewing angle value in the following for the deprojection.

Concerning the bulk Lorentz factors, the high apparent speed measured for C2 indicates that the plasma is fast and highly boosted in the regions probed at 15 GHz ($\Gamma>14$). On the other hand, the low apparent speeds and the increase of the jet opening angle towards the jet apex
suggests that the Lorentz factor is lower in the vicinity of the core than further downstream. 
For a jet orientation at $\theta_{\rm c}=(4.1 \pm 0.2)^{\circ}$ and the observed low apparent speed of A1 ($\beta_\mathrm{app}\sim 4.2$) we can estimate the jet Lorentz-factor to be of the order of 6  in the regions probed by millimeter VLBI.
This also implies that the Doppler factor $\delta=1/(\gamma\cdot(1-\beta\cos\theta))$ in the core region is not very large ($\delta\leq10$), which is in good agreement with the moderate variability of the source. 

Having measured the opening angle and the Lorentz factor at different locations along the jet, we can also test how are these two quantities related to each other. Based on hydrodynamical  \citep{1979ApJ...232...34B} and magneto-hydrodynamical models  \citep[][]{2007MNRAS.380...51K} we expect the Lorentz factor $\Gamma$ and the intrinsic jet opening angle $\phi$ to be inversely proportional, with $\Gamma\phi<1$ for a causally connected jet. This is consistent with our results. In the outer jet, the constant apparent opening angle of ${\sim}23^{\circ}$ implies an intrinsic full opening angle $\phi\sim1.6^{\circ}$, thus the product $\Gamma\phi$ yields $\sim0.4$\,$\rm rad$ for $\Gamma=14$. In the inner jet, speeds should be lower, as discussed, and the opening angle is larger. Assuming as a reference an apparent opening angle of  ${\sim}40^{\circ}$, measured at distances of $0.1-0.3$\,$\rm mas$ from the core (Fig. 6, bottom panel), and $\Gamma=6$ as estimated above for the jet base, we obtain $\Gamma\phi\sim0.30$\,$\rm rad$. Both products in the two regions are in good agreement with the median value obtained for the MOJAVE sample, $\Gamma\phi=0.35$\,$\rm rad$ \citep{2017MNRAS.468.4992P}, as well as with the results of a statistical modeling considering the same population \citep{2013A&A...558A.144C}\footnote{Note that \cite{2013A&A...558A.144C} consider the half opening angle, obtaining $\Gamma\theta=0.2$. Since we consider the full opening angle, there is a factor-of-2 difference in the product.}.

\subsection{Location of the gamma-ray emission}
\label{subs:gamma_loc}

In Section~\ref{sec:analysis_results}, we investigated the existence of correlations between the variability observed in the radio band (15\,GHz and 235\,GHz) and the $\gamma$-ray band, and we derived time lags indicating that the high-energy activity leads the one in radio (see Table~\ref{table:gamma_loc}). Having identified the most likely intrinsic parameters of the innermost jet regions, we can now translate these time lags into de-projected physical scales. Following  \cite{2011A&A...532A.146L,2010ApJ...722L...7P}, the distance between the dominant emission regions in two different bands $\Delta r$ is related to the time lag $\Delta t$ as:

\begin{equation}
\Delta r= \frac{\beta_\mathrm{app}c(\Delta t)}{\sin \theta (1+z)}.
\label{eq:delta-r}
\end{equation}

In our calculations, we adopt the mean time lags obtained in the DCCF analysis and through the Spearman's rho test analysis, the apparent speed of the innermost moving component, A1, as the best representative of the plasma speed in the region close to the core, and a viewing angle of $4.1^{\circ}$. The de-projected distances are in parsecs, for the three considered pairs of variability curves:

$$\Delta r_{\gamma-15}=(1.35 \pm 0.81)\,{\rm pc}$$
$$\Delta r_{\gamma-235}=(1.30 \pm 0.81)\,{\rm pc}$$
$$\Delta r_{235-15}=(0.34 \pm 0.29)\,{\rm pc}$$

Both the 2D cross-correlation between pairs of VLBI images and the results of the variability analysis indicate that the core shift in the frequency range between 15\,GHz - 86\,GHz
is well below the resolution limit of our observations. In fact, a de-projected shift of 0.34\,pc inferred between 235 and 15\,GHz would translate into a projected angular separation of only ${\sim}0.003$\,mas, which will be even smaller between, e.g., 43\,GHz and 86\,GHz. This result suggests that the observed VLBI cores may not be associated with the unit radio opacity surfaces ($\tau=1$), but that the core region may be a stationary shock. Negligible core shifts are derived for several blazars in the MOJAVE sample \citep{2012A&A...545A.113P} and, especially at higher frequencies, the characteristics of the VLBI core often resemble those of a stationary shock \citep[see][and references therein]{Jorstad_2017}.

Based on the analysis of the jet expansion profile presented in Sect.~\ref{subsub:apex_loc}, this possibly stationary core feature is located at a distance  $R \leq (0.019\pm0.009)$\,$\rm mas$ from the jet apex, which translates to a de-projected separation of $\leq (2.05\pm0.97)$\,pc for a viewing angle of $4.1^{\circ}$. Following the results of the variability study for the most strongly correlated pair $\gamma$-1\,mm ($\Delta r=1.30\pm0.81$\,$\rm pc$), we estimate the $\gamma$-ray emission location to be at a distance of $\sim (0.75\pm1.26)$\,pc downstream from the jet apex. By taking into account the estimated error bars, this result tells us that that the high-energy event occurs on scales ranging from sub-parsec to about ${\sim}2$\,pc distance from the jet apex.

On such scales, the most likely mechanism leading to high-energy production is highly dependent on the source type. In powerful blazars known as Flat-Spectrum Radio Quasars (FRSQs), intense external photon fields originating in the accretion disk, the Broad Line Region (BLR) or the dusty torus are likely to act as seeds for the inverse Compton radiation, while in BL Lac objects these fields are expected to be less prominent or even absent, and the high-energy emission is often well reproduced by synchrotron self-Compton models. 

As mentioned in Sect.~\ref{sec:intro}, the classification of TXS~2013$+$370 as a BL Lac object or an FSRQ is uncertain in blazar catalogs \citep[e.g.,][]{2015Ap&SS.357...75M}; this is due to the significant Galactic extinction in the source direction, which prevents a solid determination of its optical properties to be obtained. However, the recently determined, relatively high redshift \citep[$z = 0.859$,][]{2013ApJ...764..135S} is uncommon among BL Lacs \citep[see, e.g., their redshift distribution in the latest \textit{Fermi}-LAT catalog,][]{2019arXiv190510771T}. Moreover, \cite{2012ApJ...746..159K} have examined the spectral energy distribution (SED) of the source, showing it is characterized by a hard X-ray photon index and by the dominance of the inverse Compton component over the synchrotron one in power output, as usually observed in FSRQs. Indeed, reasonable fits of the SED were only obtained when including an external Compton (EC) contribution.

An estimate of the BLR size in this source can be obtained by considering the bolometric luminosity of the accretion disk $L_\mathrm{d}$, as determined by \citet{2013ApJ...764..135S}, and then following the approach of \citet{2015MNRAS.448.1060G}, which assumes a spherical BLR with radius $R_\mathrm{BLR}=10^{17}L_\mathrm{d}^{1/2}$~ cm. Through this method, we obtain a radius of the order of ${\sim} 0.07$\,pc. The analysis presented in this paper points towards distances larger than this for the location of the $\gamma$-ray emission. On scales of 1-2 parsecs from the central engine, the dusty torus is the best candidate for providing a rich seed photon field. Indeed, \citet{2012ApJ...746..159K} showed that the best fit in the SED modeling of TXS~2013$+$370 required equipartition conditions for the dominant emitting region and an external radiation field with a rather low temperature ($T_{\rm ext}{\sim}10^2\,\rm K$), thus possibly originating from cold dust. Temperatures in this range are found in the outer regions of the torus. For instance, previous studies performed in 3C~454.3 and NGC~1068 \citep{2004Natur.429...47J, 2017MNRAS.470.3283S} have derived a temperature of the order 300-600\,K at a distance of $\sim$3.5\,pc from the SMBH. Therefore, the distance that we obtain from our analysis can also be considered as a lower limit for the outer radius of the dusty torus. 

A physical scenario where the EC process is supplied by infrared photons (IR) from the dusty torus is supported by several studies of powerful blazars \citep{2012MNRAS.419.1660S,2009ApJ...704...38S}. \cite{2018MNRAS.477.4749C} showed that in the vast majority of the \textit{Fermi} FSRQs, the $\gamma$-ray emission appears to originate outside the BLR. Moreover, recent findings support a torus geometry that deviates significantly from the standard picture of a "donut"-like structure \citep{Carilli_2019,2019MNRAS.tmp.2220A,Lyu_2018,2007MNRAS.380.1172H}. The torus is likely to be rather clumpy, with polar molecular clouds providing an even richer photon field available for EC scattering.

\subsection{Transition from parabolic to conical expansion}
\label{subs:parabolic_to_conical}

In summary, we comment on the transition observed in the jet expansion profile, with the jet switching from a parabolic to a conical shape (Sect.~\ref{subsub:apex_loc}). As the resolution of radio observations increases, this phenomenon is observed in more and more jets and supports the currently most favored physical models for magnetic jet launching. These predict the jet base to be actively collimated and accelerated along an extended region, up to parsec distances from the central engine \citep[e.g.,][]{Vlahakis_2004,2007MNRAS.380...51K}. 

In TXS~2013$+$370 the transition is observed at a separation of ${\sim}0.5$\,$\rm mas$ from the jet apex, corresponding to a de-projected distance of $\sim54$\,$\rm parsecs$ for $\theta=4.1^{\circ}$. For a black hole mass of $4\times10^{8}$\,M$_\sun$ \citep{2015MNRAS.448.1060G}, we estimate that the jet collimation stops at $1.5\times10^{6}$ Schwarzschild radii from the black hole, which is of the same order as the transition distance found for M87 \citep{2012ApJ...745L..28A} and other sources in the MOJAVE sample \citep{2019arXiv190701485K}. 

This result indicates that at millimeter wavelengths, we are probing jet regions where the magnetic field is still important and that the $\gamma$-ray emission in this source is produced in the magnetically-dominated part of the jet base.

\section{Conclusions}
\label{sec:conclusions}

In this study, we gave a complete picture of the radio morphology and jet evolution of the blazar TXS~2013$+$370 during the last ten years, and we constrained the location of the $\gamma$-ray emission site. For this purpose, we employed state-of-art VLBI observations from 2~cm down to 3\,mm, as well as space-VLBI data. This unique data set allowed us to investigate the kinematic properties of the source and the jet expansion profile, from which we could derive the linear separation between the VLBI core and the jet apex. In addition, high cadence single-dish observations allowed us to monitor the flux density variability of the source, and to search for correlated activity between the radio and the $\gamma$-ray bands. The results can be summarized as follows: 

\begin{enumerate}

\item A kinematic analysis revealed the appearance of three new components, A1, N and N1. Component A1 appeared between 2005 and 2006. High-resolution 43 and 86\,GHz images since 2007 as well as at 15\,GHz after 2011 allowed us to trace its trajectory. A prominent event took place in early 2009, accompanied by a close-in-time $\gamma$-ray flare. A high-resolution, 86\,GHz image revealed the new knot N in 2010. Ultimately, space-VLBI {\em RadioAstron} data in 2012 showed the emergence of another knot, labeled as N1. The appearance of knot N1 seems to precede the enhanced emission in radio and $\gamma$-ray bands. 
    
\item The study of the jet transverse expansion profile allowed us to quantify the distance between the mm VLBI core and jet apex to be $R \leq \left( 2.05 \pm 0.97 \right)$\,pc. This analysis also revealed the existence of  a geometrical transition from a parabolic to a conical jet shape, taking place at a projected distance of $\sim 0.5$\,mas. 
This corresponds to a de-projected distance of $\sim54$\,$\rm parsecs$, or 
of ${\sim}1.5\times10^{6}$ Schwarzschild radii from the jet base.
    
\item The correlated activity between the $\gamma$-ray and radio bands enabled us to translate the observed time lags to linear distances. The strongest correlation was found between $\gamma$ rays and 1\,mm activity, with the $\gamma$-rays leading by $\left( 49 \pm 30 \right)$\,days. The estimated delay, taking into account the plasma speed on scales close to the jet apex and the viewing angle of the jet, corresponds to a de-projected distance of $\Delta r_\mathrm{\gamma-235}= \left( 1.30 \pm 0.81 \right)$\,pc. By combining the knowledge of the distance between the jet apex and the mm VLBI core and of the $\gamma$-ray production region with respect to the mm VLBI core, we constrained the location of the high-energy production site at $\left( 0.75 \pm 1.26 \right)$\,pc downstream of the jet apex. For the determined range of distances, the external seed photon field for inverse Compton emission is most likely to be produced by the dusty torus, although an origin in the BLR is also possible.

\end{enumerate}

\begin{acknowledgements}

% ##############################
I would like to thank my collaborators A.~Roy, S.~Jorstad, A.~Lobanov, V.~M.~Pati$\tilde{\mbox{n}}$o~$\acute{\mbox{A}}$lvarez, V.~Karamanavis, L.~Vega-Garc\'ia, A.-K.~Baczko, J-Y.~Kim, and I.~Myserlis, for the informative and fruitful discussions. I would also like to explicitly thank E. Angelakis for his contribution and support to this work. I thank F. Schinzel as the \textit{Fermi}-LAT internal referee, and the AGN group coordinators for the constructive comments.
I thank the anonymous referee for useful suggestions. 
I thank J.~M.~Anderson, S.~Bernhart, Y.~Y.~Kovalev, T.~Savolainen, and P.~A.~Voitsik for sharing visibility data from {\em RadioAstron}.

% ##############################

This research has made use of data obtained with the Global Millimeter VLBI Array (GMVA), which consists of telescopes operated by the MPIfR, IRAM, Onsala, Mets\"ahovi, Yebes, the Korean VLBI Network, the Green Bank Observatory and the Very Long Baseline Array (VLBA). The VLBA is a facility of the National Science Foundation operated under cooperative agreement by Associated Universities, Inc. The data were correlated at the DiFX correlator of the MPIfR in Bonn, Germany. We express our special thanks to the people supporting the observations at the telescopes during the data collection.

This research has made use of data from the OVRO 40-m monitoring program (Richards, J. L. et al. 2011, ApJS, 194, 29), which is supported in part by NASA grants NNX08AW31G, NNX11A043G, and NNX14AQ89G and NSF grants AST-0808050 and AST-1109911.

The Submillimeter Array is a joint project between the Smithsonian Astrophysical Observatory and the Academia Sinica Institute of Astronomy and Astrophysics and is funded by the Smithsonian Institution and the Academia Sinica.

This research has made use of NASA's Astrophysics Data System.

The {\em RadioAstron} project is led by the Astro Space Center of the Lebedev Physical Institute of the Russian Academy of Sciences and the Lavochkin Scientific and Production Association under a contract with the State Space Corporation ROSCOSMOS, in collaboration with partner organizations in Russia and other countries.

The \textit{Fermi}-LAT Collaboration acknowledges the generous ongoing support from a number of agencies and institutes that have supported both the development and the operation of the LAT, as well as scientific data analysis. These include the National Aeronautics and Space Administration and the Department of Energy in the United States; the Commissariat \'a l’Energie Atomique and the Centre National de la Recherche Scientifique/Institut National de Physique Nucl\'eaire et de Physique des Particules in France; the Agenzia Spaziale Italiana and the Istituto Nazionale di Fisica Nucleare in Italy; the Ministry of Education, Culture, Sports, Science and Technology (MEXT), High Energy Accelerator Research Organization (KEK), and Japan Aerospace Exploration Agency (JAXA) in Japan; and the K. A. Wallenberg Foundation, the Swedish Research  Council, and  the  Swedish  National  Space  Board in Sweden. Additional support for science analysis during the operations phase is gratefully acknowledged from the Istituto Nazionale di Astrofisica in Italy and the Centre National d'Etudes Spatiales in France. This work performed in part under DOE Contract DE-AC02-76SF00515.

\end{acknowledgements}

\bibliographystyle{aa} % or try abbrvnat or unsrtnat 
\bibliography{aanda}

\clearpage

\begin{appendix}

\section{Relevance of Interstellar scattering}
\label{ap:scattering}

TXS~2013$+$370 is located close to the Galactic plane ($b\sim 2\degr$) and in the Cygnus super-bubble region, raising the possibility that the source is affected by interstellar scattering. 
Since interstellar scattering produces a rich range of observational effects that may influence the results of the current study, we discuss it briefly. 
One manifestation induced by scattering is angular broadening, where the apparent angular size scales approximately as $\nu^{-2}$~\cite[see for example][and reference therein]{2008ApJ...672..115L}. In order to test this size-frequency relation, we combined our VLBI measurements with size measurements from \cite{2015MNRAS.452.4274P}. For the error of the angular sizes, we adopt a conservative 10\% of the FWHM.

\begin{figure}
\includegraphics[width=0.9\columnwidth]{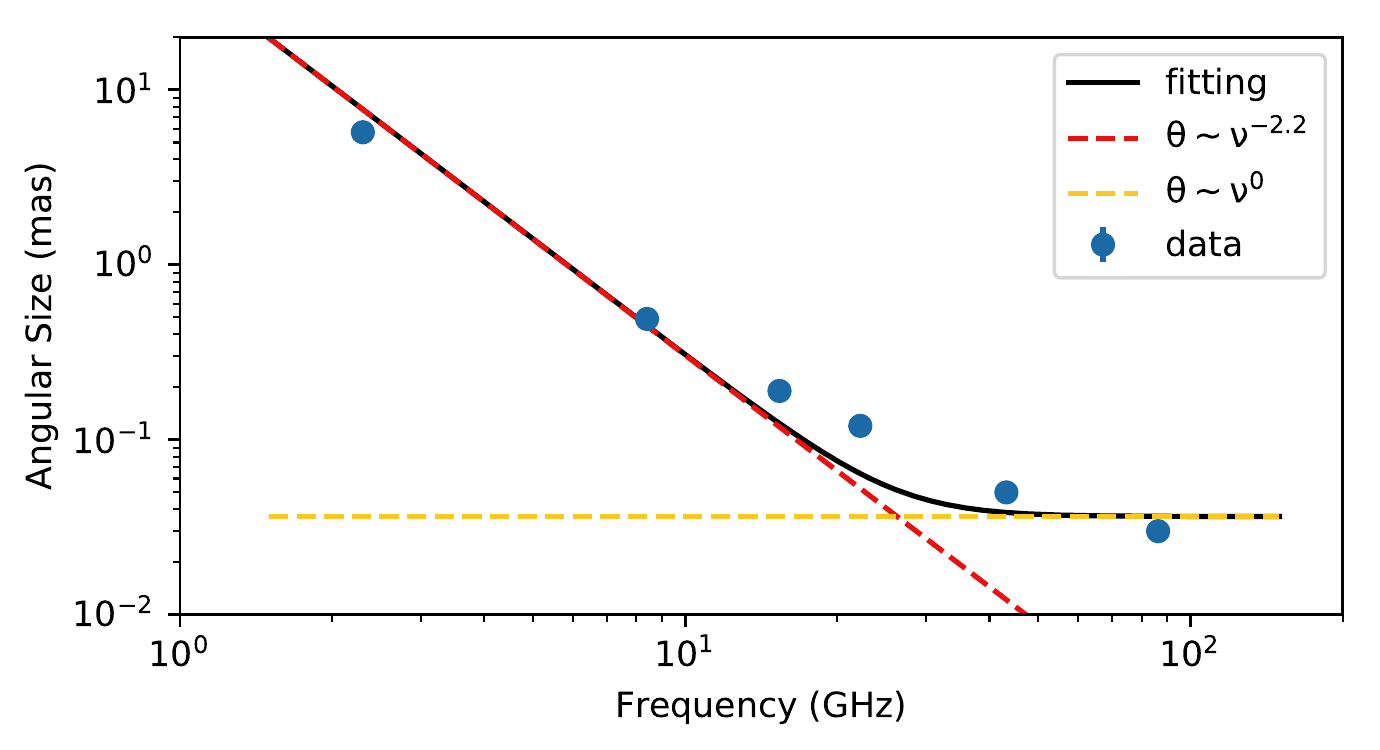} \\
\includegraphics[width=0.9\columnwidth]{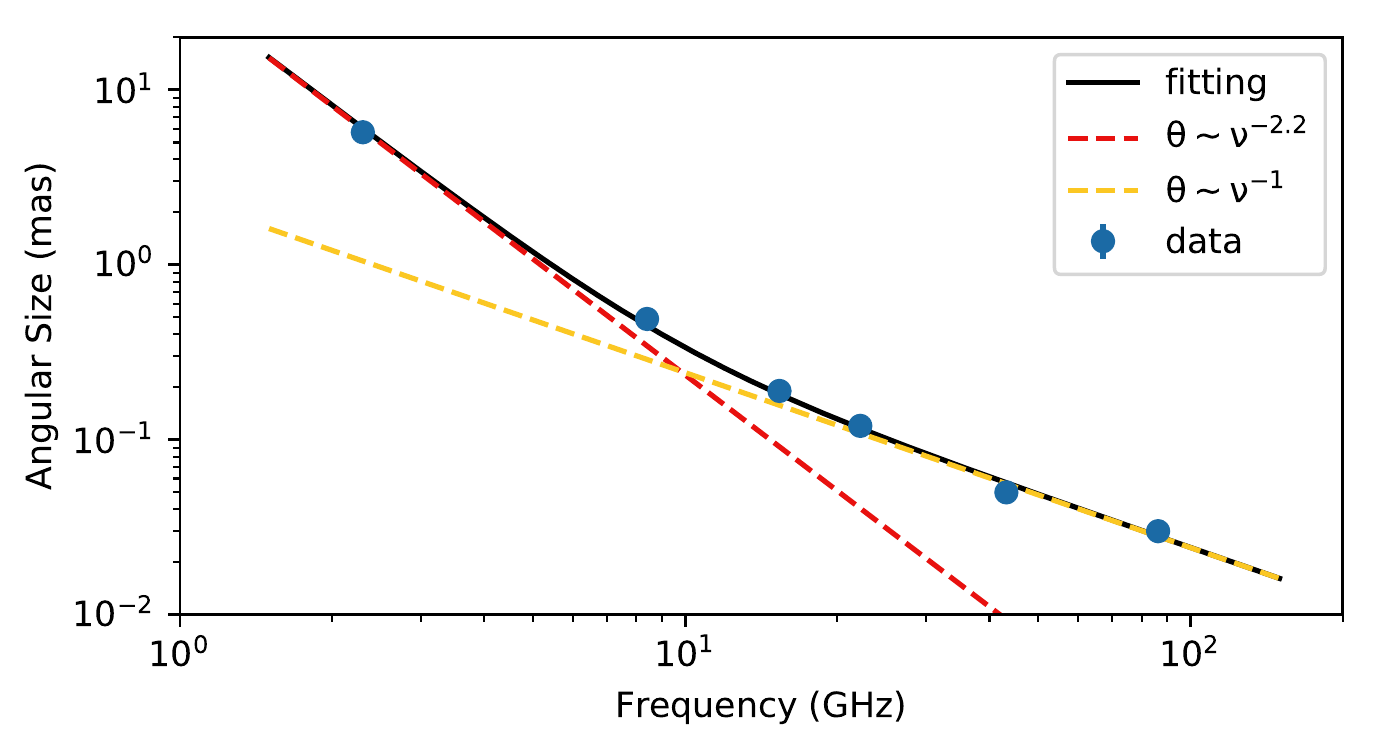}
\caption{Size-frequency relation for TXS~2013$+$370. The upper and lower panels show the case for the power-law index $k=0$ and $k=-1$, respectively. In both plots, blue circles with error bars show the measured source angular size, the solid lines indicate the best fit to the data, the red dashed line indicates the inferred scattering size, and the green dashed line indicates the inferred intrinsic size.}
\label{fig:size_freq}
\end{figure}

Following \cite{2008ApJ...672..115L}, we fit the measured angular sizes to the functional form

\begin{equation}
    \theta_{obs}^2=(\theta_{s}\nu^{-2.2})^2 + (\theta_{i}\nu^k)^2,
\label{eq:size_frequency}
\end{equation}

\noindent where $\theta_s$ and $\theta_i$ are the scattering and intrinsic source sizes, respectively. We found the best-fitting values for $\theta_s$ and $\theta_i$ using a least-squares fit approach, minimizing $\chi^2$. For the power-law index of the intrinsic size, we considered both $k=0$ (i.e., a frequency-independent intrinsic size, for a flat-spectrum radio-jet \citealt{1979ApJ...232...34B}) and $k=-1$ (i.e., a frequency scaling typical for a single inhomogenous synchrotron source \citep{1981ARA&A..19..373K}),
and selected the value of $k$ that produced the lowest $\chi^2$. The inferred scattering and intrinsic sizes from the fitting are summarized in Table~\ref{table:AB_fitting}.
In Fig.~\ref{fig:size_freq}, we plot the observed size versus frequency. The solid lines indicate the best fit of equation~(\ref{eq:size_frequency}) to the observations, and the dashed lines illustrate the 'decomposition' of the apparent source angular size into the two physical effects, scattering and intrinsic synchrotron emission. The data show that the angular broadening is prominent below 10\,GHz, which is consistent with earlier findings \citep{1986ApJ...301..312S,1989ApJ...337..730F}.
Above 10\,GHz, however, the intrinsic source size begins to dominate, where the size-frequency relation shows the expected slope for a synchrotron self-absorbed (SSA) jet.  

Another phenomenon that might be induced by interstellar scattering is the so-called `intra-day' variability (IDV, \citealt{1984AJ.....89.1111H, 1986MitAG..65..239W}), which is present in $30 -50$\,\% of all flat-spectrum quasars and BL Lac objects in cm-wavelengths \citep{1992A&A...258..279Q,lovell}. We note that TXS~2013$+$370 was observed in several IDV monitoring campaigns with the 100\,m telescope at Effelsberg at 5\,GHz, as part of the coordinated ground support for the \textit{RadioAstron} space-VLBI experiments \citep{2018Galax...6...49L}. According to $\chi^2$-tests, the source did not show significant fast variability over $\sim$ 3\,days. 
This can be explained by quenched refractive interstellar scintillation \citep{1992RSPTA.341..151N}, with a source size larger than the scattering size at 5\,GHz. The Scattering Measure (SM) in the Cygnus region varies by a factor of $2-5$ on angular scales of only a few degrees. Although TXS~2013$+$370 is scatter broadened, it shows a much lower SM than some other prominent AGN in the same region, like e.g., 2005+403, which is separated by only $\sim 4^\circ$ \cite[c.f. Fig. 6 in][]{1989ApJ...337..730F}.
We, therefore, conclude that for TXS~2013$+$370, interstellar scattering is not very strong and is dominant only at the longer cm-wavelengths. Thus it should not affect the mm-flux density and imaging significantly.

\begin{table}
\caption{Fitted scattering and intrinsic sizes and $\chi^2$. Columns from left to right: (1) Power-law index, (2) Observed angular size, (3) Intrinsic angular size, (4) Chi-square value. }    
\label{table:AB_fitting}
\centering 
\begin{tabular}{@{}cccc@{}} 
\hline\hline
$k$ & $\theta_s$ (mas) & $\theta_i$ (mas) & $\chi^2$ \\ 
\hline
0 & 48.2 $\pm$ 10.3 & (3.6 $\pm$ 1.0) $\times 10^{-2}$ &  56.9 \\
-1 & 37.1 $\pm$ 3.0 & 2.4 $\pm$ 0.1 & 3.5 \\
\hline
\end{tabular}
\end{table}

\newpage

\section{Image and model-fitting parameters.}
\label{ap:comp_param}

\begin{table*}[!htbp]
\caption{VLBI Observational Parameters. Columns from left to right: (1) Observing frequency, (2) Observing year, (3) Participating antennas, (4) Major axis of the convolving beam, (5) Minor  axis of the convolving beam, (6) Position angle, (7) Image peak flux density, (8) Noise level, (9) Image total flux density.}         
\label{table:datalog}     
\centering  
\begin{tabular}{@{}c c c c c c c c c @{}}       
\hline\hline                
Frequency  & Epoch & Array Elements & $b_\mathrm{maj} $ & $ b_\mathrm{min}$ & PA & S$_\mathrm{peak} $  & rms & $S_\mathrm{total}$ \\
(GHz) & (yyyy-mm-dd) & & (mas) & (mas) & (deg) & (Jy) & (mJy/beam) &  (Jy/beam) \\
\hline                            
15 & 2002.51 & VLBA$_{10}$ & 0.72     & 0.48     & -16.3    & 4.26     & 0.23    & 5.07\\
15 & 2003.04 & VLBA$_{10}$    & 0.77     & 0.47     & -4.89    & 2.82     & 0.08     & 3.69\\
15 & 2003.24 & VLBA$_{10}$ & 0.71     & 0.46     & -11    & 2.97     & 0.38     & 3.92\\
15 & 2005.35 & VLBA$_{10}$ & 0.75     & 0.47     & -3.62 & 2.1     & 0.39     & 2.66 \\
15 & 2005.39 & VLBA$_{10}$ & 1.10     & 0.45    & 1.18 & 2.05     & 0.21     & 2.60    \\
15 & 2005.44 & VLBA$_{10}$ & 0.99     & 0.37     & -18.1 & 1.93     & 0.54     & 2.63\\
15 & 2005.99 & VLBA$_{10}$ & 1.00     & 0.37     & -19.5 & 1.63     & 0.19     & 2.23 \\
15 & 2006.00 & VLBA$_{8}^\mathrm{ab}$ & 1.22     & 0.41     & 0.45 & 1.75     & 0.35     & 2.14\\
15 & 2006.36 & VLBA$_{10}$ & 0.77     & 0.47    & -11.2 & 1.945 & 0.19     & 1.35\\
15 & 2008.41 & VLBA$_{10}$ & 0.79     & 0.44     & -19.7 & 2.93     & 0.38     & 3.52\\
15 & 2008.75 & VLBA$_{10}$ & 0.78     & 0.45     & -14.9 & 2.56     & 0.49     & 3.07     \\
15 & 2009.15 & VLBA$_{10}$ & 0.70     & 0.45     & -16.9 & 3.36     & 0.41     & 4.3 \\
15 & 2010.46 & VLBA$_{9}^\mathrm{b}$    & 0.78  & 0.43     & -20.3 & 2.08     & 0.17     & 2.79  \\
15 & 2011.53 & VLBA$_{10}$ & 0.77     & 0.44         & -10.9    & 1.96     & 0.52     & 2.06\\
15 & 2012.48 & VLBA$_{9}^\mathrm{a}$ & 0.81     & 0.46     & -24.1    & 3.37     & 0.25     & 4.23 \\
\hline
22.2 & 2012.82 & EB+JB+RO+YS+RA     & 0.25 & 0.08 & -88.6& 1.18 & 1.81 & 3.20\\
\hline 
43 & 2007.81 & $ \mbox{VLBA}_9^\mathrm{a}$+YS+EB+ON+GB    & 0.23 & 0.1  & -23.5 & 2.75 & 0.56 & 3.46 \\
43 & 2008.79 & $ \mbox{VLBA}_9^\mathrm{a}$+YS+EB+ON+NT+GB& 0.25 & 0.1 & -15.9 & 2.17 & 0.06 & 3.77 \\
43 & 2009.21 & $ \mbox{VLBA}_{10}$+YS+EB+ON+GB & 0.26 & 0.1 & -14.6 & 4.43 & 0.33 & 6.76 \\
43 & 2009.86 & $ \mbox{VLBA}_{10}$+EB+ON+GB & 0.22 & 0.1 & -19.4 & 1.27 & 0.23 & 2.85 \\
\hline
86 & 2009.35 & $\mbox{VLBA}_8^\mathrm{c}$+PV+EB+ON+PB+MH     & 0.15 & 0.05 & -5.8 & 1.53 & 0.39 & 3.17 \\
86 & 2009.77 & $\mbox{VLBA}_8^\mathrm{c}$+PV+EB+ON+PB     & 0.14  & 0.04 & -2.56 & 1.71 & 0.83 & 3.43 \\
86 & 2010.34 & $\mbox{VLBA}_8^\mathrm{c}$+PV+EB+ON+PB+MH     & 0.18 & 0.07 & -8.72 & 1.91 & 0.36 & 3.08\\
\hline 
\end{tabular}
\tablefoot{EB: Effelsberg, JB: Jodrell Bank, RO: Radioastronomical Observatory, YS: Yebes, RA: \textit{RadioAstron}, ON: Onsala, GB: Green Bank, NT: Noto, PV: Pico Veleta, PB: Plateau de Bure, MH: Metsahovi, VLBA: Very Long Baseline Array. $^a$ : Saint Croix did not participate in the observations. $^b$ : Ovro did not participate in the observations. $^c$ : Hancock and Saint Croix do not have 3\,mm receivers.}
\end{table*}

\begin{table*}[!htbp]

\caption{Model-fitting parameters at 15\,GHz. Columns from left to right: (1) Component ID, (2) Observed epoch, (3) Observing frequency, (4) Flux density, (5) Radial distance from the core, (6) Position angle, (7) Component size}
\label{table:kinem1}
\centering
\begin{tabular}{@{}c c c c c c c@{}}
\hline\hline             
ID  & Epoch & Freq.  & S$_{v} $  & r & PA & FWHM \\
 & (years) & (GHz) & (Jy) & (mas) & ($^\mathrm{o}$)  &  (mas) \\
\hline
\\
\multirow{15}{*}{Core} &    2002.51    &    15    &    2.630    $\pm$    0.263    & -    &    -    &    0.12    $\pm$    0.01    \\
&    2003.04    &    15    &    1.71    $\pm$    0.17    &    -        &    -        &    0.11    $\pm$    0.01    \\
&    2003.24    &    15    &    2.10    $\pm$    0.21    &    -        &    -        &    0.11    $\pm$    0.01    \\
&    2005.35    &    15    &    2.31    $\pm$    0.23    &    -        &    -        &    0.19    $\pm$    0.02    \\
&    2005.39    &    15    &    2.13    $\pm$    0.21    &    -        &    -        &    0.18    $\pm$    0.02    \\
&    2005.44    &    15    &    2.09    $\pm$    0.21    &    -        &    -        &    0.20    $\pm$    0.02    \\
&    2005.99    &    15    &    1.95    $\pm$    0.20    &    -        &    -        &    0.24    $\pm$    0.02    \\
&    2006.00    &    15    &    1.91    $\pm$    0.19    &    -        &    -        &    0.19    $\pm$    0.02    \\
&    2006.36    &    15    &    0.95    $\pm$    0.01    &    -        &    -        &    0.25    $\pm$    0.03    \\
&    2008.41    &    15    &    3.32    $\pm$    0.33    &    -        &    -        &    0.22    $\pm$    0.02    \\
&    2008.75    &    15    &    2.79    $\pm$    0.28    &    -        &    -        &    0.19    $\pm$    0.02    \\
&    2009.15    &    15    &    3.81    $\pm$    0.38    &    -        &    -        &    0.23    $\pm$    0.02    \\
&    2010.46    &    15    &    1.91    $\pm$    0.19    &    -        &    -        &    0.17    $\pm$    0.02    \\
&    2011.53    &    15    &    1.43    $\pm$    0.14    &    -        &    -        &    0.16    $\pm$    0.02    \\
&    2012.48    &    15    &    3.41    $\pm$    0.34    &    -        &    -        &    0.20    $\pm$    0.02    \\
\\                                        
\multirow{17}{*}{C3}    &    2002.51    &    15    &    2.21    $\pm$    0.22    &    0.153    $\pm$    0.118    &    -156.8    $\pm$    0.7    &    0.22    $\pm$    0.02    \\
&    2003.04    &    15    &    1.78    $\pm$    0.18    &    0.27    $\pm$    0.12    &    -147.5    $\pm$    0.4    &    0.22    $\pm$    0.02    \\
&    2003.24    &    15    &    1.41    $\pm$    0.14    &    0.24    $\pm$    0.11    &    -147.0    $\pm$    1.2    &    0.18    $\pm$    0.02    \\
&    2005.35    &    15    &    0.015    $\pm$    0.002    &    0.43    $\pm$    0.10    &    -156.5    $\pm$    0.3    &    0.20    $\pm$    0.02    \\
&    2005.39    &    15    &    0.12    $\pm$    0.01    &    0.26    $\pm$    0.23    &    -169.1    $\pm$    0.7    &    0.32    $\pm$    0.03    \\
&    2005.44    &    15    &    0.18    $\pm$    0.02    &    0.26    $\pm$    0.12    &    -169.1    $\pm$    0.4    &    0.30    $\pm$    0.03    \\
&    2005.99    &    15    &    0.05    $\pm$    0.01    &    0.43    $\pm$    0.12    &    -161.2    $\pm$    0.3    &    0.20    $\pm$    0.02    \\
&    2006.00    &    15    &    0.024    $\pm$    0.002    &    0.52    $\pm$    0.14    &    -157.4    $\pm$    0.3    &    0.17    $\pm$    0.02    \\
&    2006.36    &    15    &    0.21    $\pm$    0.02    &    0.21    $\pm$    0.12    &    -158.2    $\pm$    0.5    &    0.30    $\pm$    0.03    \\
&    2008.41    &    15    &    0.11    $\pm$    0.01    &    0.26    $\pm$    0.12    &    -160.2    $\pm$    0.4    &    0.15    $\pm$    0.02    \\
&    2008.75    &    15    &    0.040    $\pm$    0.004    &    0.27    $\pm$    0.12    &    -151.1    $\pm$    0.3    &    0.19    $\pm$    0.02    \\
&    2009.15    &    15    &    0.28    $\pm$    0.03    &    0.31    $\pm$    0.11    &    -161.0    $\pm$    0.3    &    0.11    $\pm$    0.01    \\
&    2010.46    &    15    &    0.77    $\pm$    0.08    &    0.31    $\pm$    0.12    &    -155.1    $\pm$    0.4    &    0.23    $\pm$    0.02    \\
&    2011.53    &    15    &    0.82    $\pm$    0.08    &    0.21    $\pm$    0.12    &    -151.0    $\pm$    0.5    &    0.17    $\pm$    0.02    \\
&    2012.48    &    15    &    0.47    $\pm$    0.05    &    0.22    $\pm$    0.12    &    -131.4    $\pm$    0.5    &    0.16    $\pm$    0.02    \\
\\
\multirow{15}{*}{C2} &    2002.51    &    15    &    0.19    $\pm$    0.02    &    0.76    $\pm$    0.12    &    -143.0    $\pm$    0.2    &    0.42    $\pm$    0.04    \\
&    2003.04    &    15    &    0.13    $\pm$    0.01    &    0.99    $\pm$    0.12    &    -146.2    $\pm$    0.1    &    0.61    $\pm$    0.06    \\
&    2003.24    &    15    &    0.25    $\pm$    0.03    &    0.63    $\pm$    0.11    &    -142.8    $\pm$    0.7    &    0.35    $\pm$    0.04    \\
&    2005.35    &    15    &    0.24    $\pm$    0.02    &    1.17    $\pm$    0.12    &    -147.3    $\pm$    0.1    &    0.52    $\pm$    0.05    \\
&    2005.39    &    15    &    0.24    $\pm$    0.02    &    1.15    $\pm$    0.23    &    -147.1    $\pm$    0.2    &    0.50    $\pm$    0.05    \\
&    2005.44    &    15    &    0.24    $\pm$    0.02    &    1.16    $\pm$    0.12    &    -148.1    $\pm$    0.1 &    0.55    $\pm$    0.06    \\
&    2005.99    &    15    &    0.19    $\pm$    0.02    &    1.23    $\pm$    0.19    &    -148.6    $\pm$    0.1    &    0.98    $\pm$    0.10    \\
&    2006.00    &    15    &    0.14    $\pm$    0.01    &    1.29    $\pm$    0.14    &    -151.2    $\pm$    0.1    &    0.64    $\pm$    0.06    \\
&    2006.36    &    15    &    0.13    $\pm$    0.01    &    1.39    $\pm$    0.17    &    -149.7    $\pm$    0.1    &    0.83    $\pm$    0.08    \\
&    2008.41    &    15    &    0.022    $\pm$    0.002    &    2.00    $\pm$    0.12    &    -143.1    $\pm$    0.1    &    0.58    $\pm$    0.06    \\
&    2008.75    &    15    &    0.05    $\pm$    0.01    &    2.21    $\pm$    0.17    &    -146.2    $\pm$    0.1    &    0.83    $\pm$    0.08    \\
&    2009.15    &    15    &    0.026    $\pm$    0.003    &    2.19    $\pm$    0.14    &    -145.8    $\pm$    0.1    &    0.70    $\pm$    0.07    \\
&    2010.46    &    15    &    0.044    $\pm$    0.004    &    2.89    $\pm$    0.23    &    -147.4    $\pm$    0.1    &    1.17    $\pm$    0.12    \\
&    2011.53    &    15    &    0.041    $\pm$    0.004    &    3.40    $\pm$    0.22    &    -148.10    $\pm$    0.03    &    1.11    $\pm$    0.11    \\
&    2012.48    &    15    &    0.031    $\pm$    0.003    &    3.57    $\pm$    0.20    &    -147.91    $\pm$    0.03    &    0.98    $\pm$    0.10    \\
\\    
\hline                    
\end{tabular}
\end{table*}

\addtocounter{table}{-1} % make this table have the same number as the previous table
\begin{table*}[!htbp]
\caption{Model-fitting parameters at 15\,GHz. Columns from left to right: (1) Component ID, (2) Observed epoch, (3) Observing frequency, (4) Flux density, (5) Radial distance from the core, (6) Position angle, (7) Component size}
\centering
\begin{tabular}{@{}c c c c c c c@{}}
\hline\hline             
ID  & Epoch & Freq.  & S$_{v} $  & r & PA & FWHM \\
 & (years) & (GHz) & (Jy) & (mas) & ($^\mathrm{o}$)  &  (mas) \\
\hline
\\
\multirow{15}{*}{C1}    &    2002.51    &    15    &    0.11    $\pm$    0.01    &    2.77    $\pm$    0.25    &    -157.24    $\pm$    0.04    &    1.27    $\pm$    0.13    \\
&    2003.04    &    15    &    0.07    $\pm$    0.01    &    3.02    $\pm$    0.16    &    -156.00    $\pm$    0.04    &    0.82    $\pm$    0.08    \\
&    2003.24    &    15    &    0.09    $\pm$    0.01    &    3.08    $\pm$    0.28    &    -155.2    $\pm$    0.2    &    1.40    $\pm$    0.14   \\
&    2005.35    &    15    &    0.06    $\pm$    0.01    &    3.31    $\pm$    0.20    &    -156.30    $\pm$    0.03    &    1.01    $\pm$    0.10    \\
&    2005.39    &    15    &    0.06    $\pm$    0.01    &    3.38    $\pm$    0.25    &    -154.8    $\pm$    0.1    &    1.25    $\pm$    0.13    \\
&    2005.44    &    15    &    0.06    $\pm$    0.01    &    3.34    $\pm$    0.25    &    -154.28    $\pm$    0.04    &    1.23    $\pm$    0.12    \\
&    2005.99    &    15    &    0.004    $\pm$    0.004    &    3.89    $\pm$    0.16    &    -155.21    $\pm$    0.03    &    0.82    $\pm$    0.08    \\
&    2006.00    &    15    &    0.06    $\pm$    0.01    &    3.54    $\pm$    0.19    &    -154.23    $\pm$    0.04    &    0.94    $\pm$    0.09    \\
&    2006.36    &    15    &    0.06    $\pm$    0.01    &    3.40    $\pm$    0.28    &    -154.41    $\pm$    0.04    &    1.41    $\pm$    0.14    \\
&    2008.41    &    15    &    0.09    $\pm$    0.01    &    3.55    $\pm$    0.27    &    -155.74    $\pm$    0.03    &    1.34    $\pm$    0.13    \\
&    2008.75    &    15    &    0.07    $\pm$    0.01    &    3.72    $\pm$    0.23    &    -156.74    $\pm$    0.03    &    1.14    $\pm$    0.11    \\
&    2009.15    &    15    &    0.08    $\pm$    0.01    &    3.55    $\pm$    0.26    &    -156.04    $\pm$    0.03    &    1.28    $\pm$    0.13    \\
&    2010.46    &    15    &    0.06    $\pm$    0.01    &    4.15    $\pm$    0.25    &    -158.20    $\pm$    0.03    &    1.24    $\pm$    0.12    \\
&    2011.53    &    15    &    0.041    $\pm$    0.004    &    4.41    $\pm$    0.25    &    -158.85    $\pm$    0.03    &    1.24    $\pm$    0.12    \\
&    2012.48    &    15    &    0.06    $\pm$    0.01    &    4.48    $\pm$    0.26    &    -158.65    $\pm$    0.03    &    1.30    $\pm$    0.13    \\
\\
\multirow{2}{*}{A1}    &2011.53    &    15    &    0.18    $\pm$    0.02    &    0.57    $\pm$    0.12    &    -57.5    $\pm$    0.2    &    0.38    $\pm$    0.04    \\
&    2012.48    &    15    &    0.13    $\pm$    0.01    &    0.64    $\pm$    0.12    &    -159.9    $\pm$    0.2    &    0.60    $\pm$    0.06    \\    
\hline                                                    
\end{tabular}
\end{table*}

\newpage

\begin{table*}[!htbp]
\caption{Model-fitting parameters at 22\,GHz. Columns from left to right: (1) Component ID, (2) Observed epoch, (3) Observing frequency, (4) Flux density, (5) Radial distance from the core, (6) Position angle, (7) Component size}
\label{table:kinem3}
\centering
\begin{tabular}{@{}c c c c c c c@{}}
\hline\hline             
ID  & Epoch & Freq.  & S$_{v} $  & r & PA & FWHM \\
 & (years) & (GHz) & (Jy) & (mas) & ($^\mathrm{o}$)  &  (mas) \\
\hline
\\
Core &    2012.82    &    22    & 1.63    $\pm$    0.16     &    -        &    -    &    0.12     $\pm$    0.01     \\
N1 &    2012.82    &    22    &    0.90    $\pm$    0.09    &    0.11    $\pm$    0.03    &    -81.9    $\pm$    0.3    &    0.09    $\pm$    0.01    \\
C3 &    2012.82    &    22    &    0.410    $\pm$    0.004    &    0.26    $\pm$    0.03    &    -121.9    $\pm$    0.1    &    0.07    $\pm$    0.01    \\
A1 &    2012.82    &    22    &    0.19    $\pm$    0.02    &    0.57    $\pm$    0.06    &    -133.7    $\pm$    0.1    &    0.29    $\pm$    0.03    \\
N &    2012.82    &    22    &    0.11    $\pm$    0.01    &    0.71    $\pm$    0.03      &    -153.4    $\pm$    0.04    &    0.17    $\pm$    0.02    \\
\\
\hline
\end{tabular}
\end{table*}

\begin{table*}[!htbp]
\caption{Model-fitting parameters at 43\,GHz. Columns from left to right: (1) Component ID, (2) Observed epoch, (3) Observing frequency, (4) Flux density, (5) Radial distance from the core, (6) Position angle, (7) Component size}
\label{table:kinem4}
\centering
\begin{tabular}{@{}c c c c c c c@{}}
\hline\hline             
ID  & Epoch & Freq.  & S$_{v} $  & r & PA & FWHM \\
 & (years) & (GHz) & (Jy) & (mas) & ($^\mathrm{o}$)  &  (mas) \\
\hline
\\
\multirow{3}{*}{Core} &    2007.81    &    43    &    3.05    $\pm$    0.31    &    -        &    -    &    0.05    $\pm$    0.01    \\
&    2008.79    &    43    &    2.46    $\pm$    0.25    &    -    &    -    &    0.05    $\pm$    0.01    \\
&    2009.21    &    43    &    5.00    $\pm$    0.50    &    -    &    -    &    0.05    $\pm$    0.01    \\
&    2009.86    &    43    &    1.38    $\pm$    0.13    &    -    &    -    &    0.05    $\pm$    0.01    \\
\\
\multirow{4}{*}{A2} &    2007.81    &    43    &    0.023    $\pm$    0.002    &    0.11    $\pm$    0.08    &    -103.0    $\pm$    0.3    &    0.023    $\pm$    0.002    \\
&    2008.79    &    43    &    0.015    $\pm$    0.002    &    0.12    $\pm$    0.08    &    -104.0    $\pm$    0.3    &    0.06    $\pm$    0.01    \\
&    2009.21    &    43    &    0.25    $\pm$    0.03    &    0.12    $\pm$    0.08    &    -119.0    $\pm$    0.3    &    0.06    $\pm$    0.01    \\
&    2009.86    &    43    &    0.95    $\pm$    0.09    &    0.14    $\pm$    0.03      &    -123.9    $\pm$    0.2    &    0.10    $\pm$    0.01    \\
\\
\multirow{3}{*}{A1}    &    2007.81    &    43    &    0.23    $\pm$    0.02    &    0.17    $\pm$    0.08    &    -172.6    $\pm$    0.2    &    0.09    $\pm$    0.01    \\
&    2008.79    &    43    &    1.19    $\pm$    0.12    &    0.21    $\pm$    0.08    &    -152.9    $\pm$    0.1    &    0.13    $\pm$    0.01    \\
&    2009.21    &    43    &    1.51    $\pm$    0.15    &    0.24    $\pm$    0.08    &    -157.1    $\pm$    0.1    &    0.13    $\pm$    0.01    \\
&    2009.86    &    43    &    0.42    $\pm$    0.04    &    0.30    $\pm$    0.05    &    -154.6    $\pm$    0.11    &    0.20    $\pm$    0.02    \\
\\
\multirow{3}{*}{C3}    &    2007.81    &    43    &    0.10    $\pm$    0.01    &    0.32    $\pm$    0.08    &    -160.5    $\pm$    0.1    &    0.22    $\pm$    0.02    \\
&    2008.79    &    43    &    0.033    $\pm$    0.003    &    0.37    $\pm$    0.03    &    -166.4    $\pm$    0.1    &    0.038    $\pm$    0.004    \\
&    2009.21    &    43    &    0.020    $\pm$    0.002    &    0.37    $\pm$    0.03    &    -164.2    $\pm$    0.1    &    0.027    $\pm$    0.003    \\
&    2009.86    &    43    &    0.07    $\pm$    0.01    &    0.46    $\pm$    0.05    &    -163.4    $\pm$    0.1    &    0.14    $\pm$    0.01    \\
\hline
\end{tabular}
\end{table*}

%\vspace{2cm}

\begin{table*}[!htbp]
\caption{Model-fitting parameters at 86\,GHz. Columns from left to right: (1) Component ID, (2) Observed epoch, (3) Observing frequency, (4) Flux density, (5) Radial distance from the core, (6) Position angle, (7) Component size}. 
\label{table:kinem5}
\centering
\begin{tabular}{@{}c c c c c c c@{}}
\hline\hline             
ID  & Epoch & Freq.  & S$_{v} $  & r & PA & FWHM \\
 & (years) & (GHz) & (Jy) & (mas) & ($^\mathrm{o}$)  &  (mas) \\
\hline
\multirow{3}{*}{Core} &    2009.35    &    86    &    2.04    $\pm$    0.20    &    -        &    -    &    0.037    $\pm$    0.004    \\
&    2009.77    &    86    &    1.58    $\pm$    0.16    &    -    &    -    &    0.013    $\pm$    0.001    \\
&    2010.34    &    86    &    2.07    $\pm$    0.21    &    -    &    -    &    0.030    $\pm$    0.003    \\
\\
\multirow{4}{*}{A2}    &    2009.35    &    86    &    0.31    $\pm$    0.03    &    0.08    $\pm$    0.04    &    -117.9    $\pm$    0.2    &    0.05    $\pm$    0.01    \\
&    2009.77    &    86    &    1.72    $\pm$    0.17    &    0.09    $\pm$    0.04    &    -125.9    $\pm$    0.2    &    0.012    $\pm$    0.001    \\
& 2010.34    &    86    &    0.072    $\pm$    0.002    &    0.06    $\pm$    0.06    &    -119.9    $\pm$    0.1    &    0.037    $\pm$    0.002 \\
\\
\multirow{2}{*}{A1}    &    2009.35    &    86    &    0.83    $\pm$    0.08    &    0.25    $\pm$    0.09    &    -159.8    $\pm$    0.1    &    0.18    $\pm$    0.02    \\
&    2009.77    &    86    &    0.25    $\pm$    0.03    &    0.30    $\pm$    0.05    &    -171.6    $\pm$    0.1    &    0.10    $\pm$    0.01    \\
\\
A1$+$C3 &    2010.34    &    86    &    0.29    $\pm$    0.03    &    0.40    $\pm$    0.09    &    -155.7    $\pm$    0.1    &    0.17    $\pm$    0.02    \\
\\
\multirow{1}{*}{N} & 2010.34    &    86    &    0.71    $\pm$    0.07    &    0.16    $\pm$    0.08    &    -134.5    $\pm$    0.1    &    0.16    $\pm$    0.02 \\
\hline                    
\end{tabular}
\end{table*}

\end{appendix}

%-------------------------------------------------------------------

\end{document}